%% file: ms.tex
\documentclass[reprint,prmaterials,aps,superscriptaddress,showkeys]{revtex4-2}

\usepackage[T1]{fontenc}       
\usepackage{xcolor}
\usepackage[version=4]{mhchem}
\usepackage{booktabs}          
\usepackage{longtable}
\usepackage{graphicx}
\usepackage{hyperref}

\hypersetup{
  pdftitle={Quantifying uncertainty in high-throughput density functional theory: a
  comparison of AFLOW, Materials Project, and OQMD},
  pdfauthor={Vinay I. Hegde <vhegde@citrine.io>},
  colorlinks=true,
  linkcolor=blue,
  citecolor=blue,
  urlcolor=blue,
}

\begin{document}
\title{Quantifying uncertainty in high-throughput density functional theory: a
comparison of AFLOW, Materials Project, and OQMD}

\author{Vinay I.\ Hegde}\thanks{These authors contributed equally to this work}
\author{Christopher K.\ H.\ Borg}\thanks{These authors contributed equally to
this work}
\affiliation{Citrine Informatics, 2629 Broadway, Redwood City, CA 94063}
\author{Zachary del Rosario}
\affiliation{Citrine Informatics, 2629 Broadway, Redwood City, CA 94063}
\affiliation{Olin College of Engineering, 1000 Olin Way, Needham, MA 02492}
\author{Yoolhee Kim}
\author{Maxwell Hutchinson}
\author{Erin Antono}
\author{Julia Ling}
\affiliation{Citrine Informatics, 2629 Broadway, Redwood City, CA 94063}
\author{Paul Saxe}
\affiliation{Molecular Sciences Software Institute, Virginia Tech, Blacksburg,
VA 24061}
\author{James E.\ Saal}
\author{Bryce Meredig}
\email{bryce@citrine.io}
\affiliation{Citrine Informatics, 2629 Broadway, Redwood City, CA 94063}

\date{\today}

\begin{abstract}
\input{sections/abstract}
\end{abstract}


\maketitle

\section{Introduction}\label{sec:intro}
\input{sections/introduction}

\section{Methods}\label{sec:methods}
\input{sections/methods}

\section{Results}\label{sec:results}
\input{sections/results}

\section{Discussion}\label{sec:discussion}
\input{sections/discussion}

\section{Conclusion}\label{sec:conclusion}
\input{sections/conclusion}

\section*{Conflicts of interest}\label{sec:coi}
ZdR was previously employed by Citrine Informatics.
PS has worked as a subcontractor to Citrine Informatics.

\section*{Acknowledgements}\label{sec:ack}
This material is based upon work supported by the U.S.\ Department of Energy,
Office of Science, Office of Basic Energy Sciences, Small Business Technology
Transfer Program under Award Number DE-SC0015106.
The authors would like to thank Cormac Toher for advice on using the AFLUX
RESTful API, and Anubhav Jain, Shyue Ping Ong, Matthew Horton, Eric B.\ Isaacs,
and Chris Wolverton for their comments on an earlier version of this
manuscript.

\section*{Author Contributions}\label{sec:author_contrib}
Conceptualization: C.K.H.B., V.H., P.S., M.H., J.E.S., B.M.;
Methodology: C.K.H.B., V.H., E.A., Y.K., M.H., P.S., J.E.S.;
Software: C.K.H.B., V.H., E.A., Y.K.;
Validation: C.K.H.B., V.H., B.M.;
Formal analysis: C.K.H.B., V.H., Z.d.R., E.A., Y.K.;
Investigation: C.K.H.B., V.H.;
Data Curation: C.K.H.B., V.H.;
Writing -- Original Draft: C.K.H.B., V.I.H., M.H., P.S., J.E.S.;
Writing -- Review \& Editing: all authors;
Visualization: E.A., Y.K., C.K.H.B., V.H.;
Supervision: J.E.S., B.M., J.L.

\section*{Data Availability}\label{sec:data_avail}
All data and Python scripts required to perform the analysis presented in this
work are made available via the GitHub repository at
\url{https://github.com/CitrineInformatics-ERD-public/htdft-uq}.

\clearpage

\appendix

\section{Definitions of statistical quantities}\label{sec:metric_defs}
\input{sections/metric_defs}

\bibliography{references}

\end{document}


\title{Supplemental Information for ``Quantifying uncertainty in
high-throughput density functional theory: a comparison of AFLOW, Materials
Project, and OQMD''}

\author{Vinay I.\ Hegde}\thanks{These authors contributed equally to this work}
\author{Christopher K.\ H.\ Borg}\thanks{These authors contributed equally to
this work}
\affiliation{Citrine Informatics, 2629 Broadway, Redwood City, CA 94063}
\author{Zachary del Rosario}
\affiliation{Citrine Informatics, 2629 Broadway, Redwood City, CA 94063}
\affiliation{Olin College of Engineering, 1000 Olin Way, Needham, MA 02492}
\author{Yoolhee Kim}
\author{Maxwell Hutchinson}
\author{Erin Antono}
\author{Julia Ling}
\affiliation{Citrine Informatics, 2629 Broadway, Redwood City, CA 94063}
\author{Paul Saxe}
\affiliation{Molecular Sciences Software Institute, Virginia Tech, Blacksburg,
VA 24061}
\author{James E.\ Saal}
\author{Bryce Meredig}
\email{bryce@citrine.io}
\affiliation{Citrine Informatics, 2629 Broadway, Redwood City, CA 94063}
\date{\today}

\maketitle

\onecolumngrid

\section{HT-DFT Databases and Management Codes}\label{sec:htdft_tables}
\input{sections/htdft_tables}
\clearpage

\section{Data Management}\label{sec:data_management}
\input{sections/data_management}
\clearpage

\section{Distribution of compounds in formation energy
bimodal}\label{sec:dhf_bimodal}
\input{sections/dhf_bimodal}
\clearpage

\section{Per-material-class median absolute differences}\label{sec:pmc-mad}
\input{sections/pmc_mad}
\clearpage

\section{Element-wise analysis of HT-DFT differences}\label{sec:elem-wise}
\input{sections/element_wise}
\clearpage

\section{Results from the larger, multiple ICSD ID comparison
dataset}\label{sec:multi-icsd-id-results}
\input{sections/multi_icsd_id_results}
\clearpage

\section{Example Physical Information Files (PIFs)}\label{sec:example_pifs}
\input{sections/example_pifs}
\clearpage

\bibliography{references}

%% file: sections/abstract.tex
A central challenge in high throughput density functional theory (HT-DFT)
calculations is selecting a combination of input parameters and post-processing
techniques that can be used across all materials classes, while also managing
accuracy-cost tradeoffs.
To investigate the effects of these parameter choices, we consolidate three
large HT-DFT databases: Automatic-FLOW (AFLOW), the Materials Project (MP), and
the Open Quantum Materials Database (OQMD), and compare reported properties
across each pair of databases for materials calculated using the same initial
crystal structure.
We find that HT-DFT formation energies and volumes are generally more
reproducible than band gaps and total magnetizations; for instance, a notable
fraction of records disagree on whether a material is metallic (up to 7\%) or
magnetic (up to 15\%).
The variance between calculated properties is as high as 0.105~eV/atom (median
relative absolute difference, or MRAD, of 6\%) for formation energy,
0.65~{\AA}$^3$/atom (MRAD of 4\%) for volume, 0.21~eV (MRAD of 9\%) for band
gap, and 0.15~$\mu_{\rm B}$/formula unit (MRAD of 8\%) for total magnetization,
comparable to the differences between DFT and experiment.
We trace some of the larger discrepancies to choices involving
pseudopotentials, the DFT+\textit{U} formalism, and elemental reference states,
and argue that further standardization of HT-DFT would be beneficial to
reproducibility.

%% file: sections/introduction.tex
Over the past decade, high-throughput (HT) density functional theory (DFT) has
emerged as a widely-used tool for materials discovery and
design~\cite{curtarolo2013high, jain2016computational, saal2013materials}.
In a standard HT-DFT workflow, software tools automate the process of
calculating materials properties of interest within DFT, including submitting
jobs to high-performance computing infrastructure, on-the-fly error handling,
post-processing and dissemination of results, and so on, enabling researchers
to evaluate typically 10$^3$--10$^6$ materials with minimal human intervention.
The resulting database can then be screened for candidate materials exhibiting
promising combinations of calculated properties or to search for trends amongst
materials behavior to gain new chemical insights or develop surrogate models.

The increasingly widespread usage of HT-DFT in materials research can be
attributed to a combination of three key factors.
First, a large number of specialized codes implement fully automated
calculations of specific materials properties within DFT, ranging from phonon
dispersions to dielectric tensors.
For example, VASP 5.1~\cite{vasp-1, vasp-2} introduced a feature enabling users
to calculate elastic tensors by simply setting a parameter in the input file.
Second, the ongoing growth of computing power has ensured that HT-DFT is now
well within reach of a single university research group.
Third, sophisticated, free, often open-source, software is readily available
for managing large numbers of DFT calculations, post-processing output, and
storing the resulting data systematically in databases.
Thus, a number of HT-DFT databases with various focus areas have emerged; a
list of exemplars, including any supporting workflow automation software, is
given in Section~S-I of the Supplemental Information (SI).

However, the entirely-automated nature of HT-DFT introduces a few key
challenges.
First, by definition, the volume of data from HT-DFT is too high for each
individual calculation to undergo manual review or
analysis~\cite{curtarolo2013high}.
How, then, are the quality and integrity of calculations monitored in
high-throughput?
Second, HT-DFT requires choosing, often at the outset, settings that are
consistent across all calculations, encompassing all materials classes and
properties being calculated.
For example, it may not be known \textit{a priori} whether the material being
calculated is a metal or an insulator.
As a result, the calculation parameters that affect, e.g., how
electronic occupancies are smeared near the Fermi level must be chosen so that
they are applicable to both metals and insulators.
Third, practical HT-DFT calculations involve balancing accuracy
and computational cost; best-practice
recommendations~\cite{mattsson2004designing} involve steps such as explicit
convergence tests, which become computationally infeasible in the HT context.
Of these challenges, only the first, related to monitoring the quality and
integrity of calculations in high-throughput has been addressed.
Software frameworks, such as Custodian~\cite{ong2020custodian},
qmpy~\cite{kirklin2015open}, and AiiDa~\cite{huber2020aiida}, can store
provenance information to ensure the integrity of calculations, and gracefully
handle errors associated with catastrophic failures, e.g., those related to
file read/write operations or memory issues during runtime, insufficient
walltimes on high-performance computing resources, and misconfiguration of the
underlying numerical libraries.

Since HT-DFT has become increasingly central to materials informatics efforts
across the spectrum, from high-throughput screening to machine
learning~\cite{zhuo2018identifying, meredig2014combinatorial} it is crucial to
resolve the following concerns:
(a) There is no one ``correct'' solution to some of the challenges of HT-DFT
mentioned above, and different databases have tackled them slightly
differently.
How sensitive are the calculated materials properties to the different HT-DFT
parameter choices?
(b) The focus areas of many prominent HT-DFT databases in terms of the
materials and properties calculated are often quite different.
As a result, materials data from the various HT-DFT databases are often mixed
with one another for thermochemical or other analysis.
How interoperable are these various calculated materials properties across
HT-DFT databases?
We emphasize that such a comparison across HT-DFT databases is different from
analyzing the reproducibility of DFT across software implementations and
potentials, e.g. focusing on equations of state of elemental
crystals:~\cite{lejaeghere2016reproducibility} the challenges of HT-DFT lie in
choosing parameters that are applicable across a wide variety of materials and
properties, targeting both reasonable accuracy \textit{and} computational
cost---very distinct from performing highly-accurate DFT calculations of a
small set of materials.

Here, we analyze the reproducibility and interoperability of HT-DFT
calculations.
We critically compare the agreement between three databases for four
properties: formation energy ($\Delta E_{\rm f}$), volume ($V$), band gap
($E_{\rm g}$), and total magnetization ($M$).
We find certain properties (formation energies and volumes) to be more
consistent across databases than others (band gap and magnetization).
We then quantify the variability in each of the properties across databases and
find that the typical differences between two HT-DFT databases are similar to
those between DFT and experiment.
Finally, we compare properties across different materials classes to identify
characteristics of materials and/or properties that are harder than others to
reproduce.
In all cases, we identify trends, surface outliers, and investigate potential
causes for an observed systematic differences between the databases.

%% file: sections/methods.tex
We focus on three prominent HT-DFT databases in this work:
Automatic FLOW (AFLOW)~\cite{curtarolo2012aflowlib}, the Materials Project
(MP)~\cite{jain2013commentary}, and the Open Quantum Materials Database
(OQMD)~\cite{saal2013materials, kirklin2015open}.
All three databases contain calculations of a large number of
mostly-experimentally reported, ordered compounds from the Inorganic Crystal
Structure Database (ICSD)~\cite{belsky2002new}.
In addition, they contain calculations of many thousands of hypothetical
compounds generated from common structural prototypes or other informatics
approaches.
As noted earlier, there are many other large HT-DFT databases, e.g.,
JARVIS-DFT~\cite{choudhary2020joint}, Materials
Cloud~\cite{talirz2020materials}, and others listed in Table S-I of the SI.
Here, we limit our focus to AFLOW, Materials Project, and OQMD as the latter
(a) are among the longest-running, mature, widely-used, and general-purpose,
and (b) use the VASP software package~\cite{vasp-1, vasp-2} and projector
augmented wave (PAW) potentials~\cite{Blochl1994, Kresse1999} with the
Perdew-Burke-Ernzerhof (PBE) parameterization~\cite{Perdew1996} of a
generalized-gradient approximation (GGA) to the DFT exchange-correlation
functional.
The variance in HT-DFT-calculated properties studied in the present work is,
therefore, almost entirely due to differences in various choices involved in
HT-DFT (e.g., those involving calculation parameters such as $k$-point density,
the DFT+$U$ approach, post-calculation processing techniques, different
versions of VASP and any associated software bugs, different versions of PBE
pseudopotentials used) and \textit{not} due to different implementations of DFT
or approximations to the underlying exchange-correlation functional itself.

AFLOW has standardized band structure calculations~\cite{setyawan2010high,
setyawan2011scintillators}, binary alloy cluster
expansions~\cite{levy2010uncovering}, finite-temperature thermodynamic
properties~\cite{toher2014high}, elastic and thermomechanical
properties~\cite{toher2017combining} calculated for many materials, and has an
application programming interface (API) based on the REpresentational State
Transfer (REST) standard (commonly referred to as ``RESTful API'') for
accessing data~\cite{taylor2014restful}.
The Materials Project includes a variety of properties calculated for specific
subsets of materials in the database, including elastic~\cite{de2015charting},
thermoelectric~\cite{chen2016understanding},
piezoelectric~\cite{de2015database}, dielectric~\cite{petousis2017high},
vibrational~\cite{petretto2018high} properties, and X-ray adsorption
spectra~\cite{zheng2018automated}.
It also includes a collection of apps such as a Pourbaix diagram
calculator~\cite{persson2012prediction}, and the underlying data are accessible
via a RESTful API~\cite{ong2015materials}.
Finally, the Open Quantum Materials Database (OQMD) contains calculations of a
large number of hypothetical compounds based on structural
prototypes,~\cite{kirklin2016high, emery2016high, wang2018crystal} and provides
tools for the construction of DFT ground state phase diagrams at ambient and
high-pressures~\cite{r2007first, hegde2020, amsler2018exploring}.
The OQMD provides the entirety of the underlying database to download all at
once, and a RESTful API for programmatic access~\cite{OQMD_API}.
License and access information for the three databases is included in Section
S-II of the SI.

We query all three databases (AFLOW: queried June 2021; MP: v2019.05; OQMD:
v1.2) for the calculated properties of materials whose crystal structures were
sourced from the ICSD and aggregate them into a single dataset, after
converting records from all sources into a unified, consistent data format, the
Physical Information File (PIF)~\cite{michel2016beyond}.
We then generate a set of comparable records for each pairwise combination of
the databases---all calculations using the same initial crystal structure, by
matching their ICSD Collection Codes (hereafter referred to as ``ICSD ID'').
In instances where more than one calculation within a single database was
labeled with the same ICSD ID, we use the lowest energy calculation for all
analysis.
In addition, we discard records with obviously unphysical property values
(those with formation energy outside the [$-5$~eV/atom, $+5$~eV/atom] window
and volumes above 150~{\AA}$^3$/atom), and normalize properties to the same
units, where required.
We then perform statistical analysis on the final curated set of comparable
records across the three databases.
Definitions of the metrics used in our analysis are given in
Appendix~\ref{sec:metric_defs} and details of the query and curation steps are
provided in Section~S-II of the SI.

%% file: sections/results.tex

\begin{table*}[!ht]
\centering
\input{tables/records_tally}
\caption{The number of records after establishing ICSD ID equivalency for each
property of interest in the AFLOW, Materials Project (MP), and OQMD HT-DFT
databases, as well as for pairwise comparisons of the three
databases.}\label{tbl:prop_table}
\end{table*}

The aggregation and processing of the data from the three HT-DFT databases
results in a set of $\sim$70,000 total comparable DFT calculations.
For each property of interest, i.e., formation energy per atom, volume per
atom, band gap, total magnetization per formula unit (f.u.), the counts of
records, and overlapping records for each pair of databases are shown in
Table~\ref{tbl:prop_table}.
Approximately 15,000--25,000 comparisons can be made for each property and
database pair, except for comparisons to formation energies from AFLOW, where
only $\sim$2,200 records are reported.
As mentioned earlier, overlapping records across databases were determined by
using exact ICSD ID matches for the reported calculations.


\subsection{Overall pairwise comparison
statistics}\label{ssec:results-overall-stats}

\begin{table*}[!ht]
\centering
\input{tables/corr_coeffs}
\caption{Overall statistics (median absolute difference (MAD), interquartile
  range (IQR), Pearson's linear correlation coefficient ($r$), and Spearman's
  rank correlation coefficient ($\rho$)) for the comparison of properties
  across HT-DFT databases.
For each property, records overlapping across a pair of databases are compared
(* for band gap and magnetization, only non-zero values are compared).
Generally, lower MAD, lower IQR, higher $r$, and higher $\rho$ values indicate
better reproducibility of calculated properties.}\label{tbl:summary_table}
\end{table*}

Table~\ref{tbl:summary_table} shows some overall statistics for comparisons of
all properties across comparable records in the three databases: the median
absolute difference (MAD), the interquartile range (IQR), the Pearson
correlation coefficient ($r$), and Spearman's rank correlation coefficient
($\rho$) (definitions of the metrics are in Appendix~\ref{sec:metric_defs}).
For band gap and total magnetization, the statistics were calculated only on
subsets of overlapping records where both databases agreed that a material is
non-metallic ($E_{\rm g} > 0.01$~eV) and is magnetic ($M > 0.01$~$\mu_{\rm
B}$/atom), respectively.
The latter threshold on the per-formula unit total magnetization ensures that
undesired comparisons of different magnetic configurations for the same crystal
structure (i.e., ferromagnetic configuration in one database being compared to
antiferromagnetic configuration in another) are avoided as much as possible.

Overall, we find that:
(a) The MAD in formation energy across pairs of databases can be up to
0.105~eV/atom, comparable to the $\sim$0.1~eV/atom difference between DFT and
experimental formation energies~\cite{kirklin2015open}.
(b) The MAD in volume across pairs of databases can be up to
0.65~{\AA}$^3$/atom (median absolute difference relative to mean (MRAD), of
  3.8\%), comparable to error between DFT and
    experiment~\cite{haas2009calculation}.
(c) The MAD in band gap across pairs of databases can be up to 0.21~eV, even
when comparing only records where both databases agree that a material is not
metallic.
For around 5\%--7\% of overlapping records, databases disagree whether a
material is metallic.
(d) The comparison of total magnetization shows high variability across
database pairs.
While the dispersion of differences for the MP-OQMD comparison is very small
(MAD of 0.01~$\mu_{\rm B}$/f.u.\ and IQR of 0.05~$\mu_{\rm B}$/f.u.), the
dispersion of differences in comparisons with AFLOW are rather large (up to MAD
of 0.15~$\mu_{\rm B}$/f.u.\ and IQR of up to 2.0~$\mu_{\rm B}$/f.u.).
In all cases, the correlation between calculated values is lower than for the
other three properties, with both Pearson and Spearman correlation coefficients
ranging from 0.6--0.8.
We further note that the latter poor correlation exists even after excluding
overlapping records where the two databases disagree on whether the material is
magnetic (10\%--15\% of the records).


\subsection{Distribution of differences in calculated
properties}\label{ssec:results-diff-dist}

\begin{figure*}[!ht]
\centering
\includegraphics[width=0.9\textwidth]{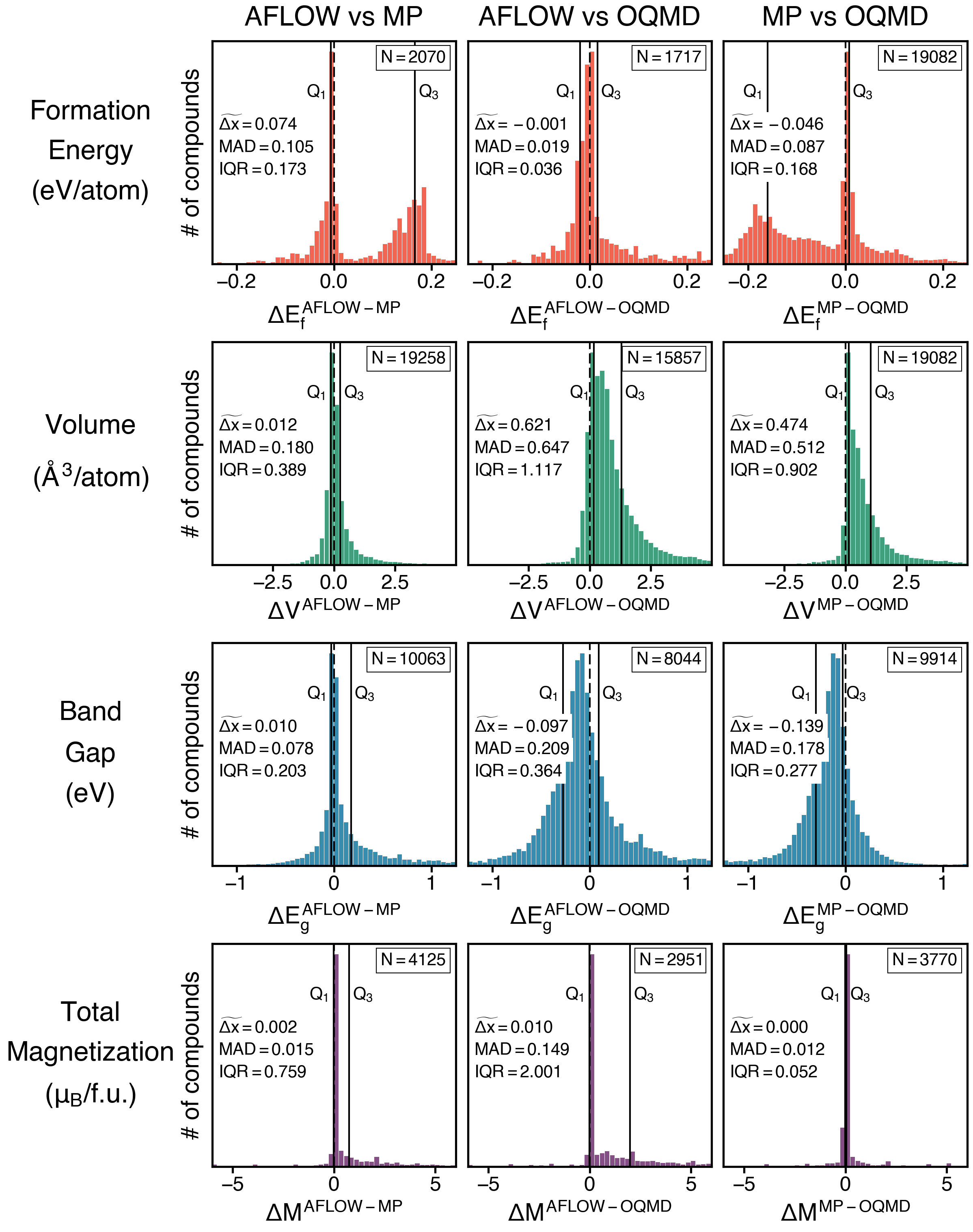}
\caption{Distribution of the differences in calculated properties across HT-DFT
  databases.
Each panel corresponds to a property and pair of databases being compared.
Solid vertical black lines correspond to the first (Q$_1$) and third (Q$_3$)
quartiles of the distribution.
The number of records overlapping across the two databases is shown in the top
right corner of each panel;
the median of distribution ($\widetilde{\Delta x}$), the median absolute
difference (MAD), and the interquartile range (IQR) are noted on the
left.}\label{fig:pair_histograms}
\end{figure*}

We first analyze the raw differences in the calculated properties for records
overlapping across pairs of databases.
Figure~\ref{fig:pair_histograms} shows the distribution of the differences in
calculated values for each of formation energy, volume, band gap, and total
magnetization, for each pairwise combination of databases.

\noindent
\textit{Formation energy}:
The distribution of differences in calculated formation energy across AFLOW-MP
and MP-OQMD is surprisingly bimodal, with peaks around 0 and $\pm0.2$~eV/atom.
We find that the peak near 0.2~eV/atom in both pairwise comparisons corresponds
mostly to oxides (see Figure~S1), and is a result of different approaches in
the two databases toward correcting DFT-calculated formation energies (see
Section~\ref{ssec:chem-pot}).
While the median difference ($\widetilde{\Delta x}$ in
Figure~\ref{fig:pair_histograms}) are reasonably small across all three
pairwise comparisons (up to $\sim$0.074~eV/atom), the difference distributions
for AFLOW-MP and MP-OQMD are rather wide.
The median absolute difference (MAD) and the interquartile range (IQR), both
robust measures of the spread of a distribution, are up to $\sim$0.105~eV/atom
and $\sim$0.173~eV/atom, respectively.

\noindent
\textit{Volume}:
The distribution of differences in calculated volumes is skewed towards smaller
volumes in the OQMD, but such a skew is absent in the AFLOW-MP comparison.
Correspondingly, the median difference between AFLOW and MP volumes are
$\sim$0.01~{\AA}$^3$/atom, whereas the median differences are
$\sim$0.62~{\AA}$^3$/atom and $\sim$0.47~{\AA}$^3$/atom for AFLOW-OQMD and
MP-OQMD, respectively.
The consistently smaller volumes calculated in the OQMD can be understood to
result from the choice of the plane wave energy cutoff used for DFT relaxation
calculations.
The OQMD chooses a plane wave cutoff that is lower than that used in AFLOW and
MP (\textsf{ENMAX} in the POTCAR file, up to 400~eV in OQMD, as opposed to
520~eV in MP and up to 560~eV in AFLOW) for full cell relaxations.
The lower plane wave cutoff results in Pulay stresses and generally smaller
volumes than fully relaxed calculations.
The MAD in volumes for comparisons, especially for OQMD with the other two
databases, is up to $\sim$0.65~{\AA}$^3$/atom.
In addition, some differences in reported volumes can result from the different
relaxation schemes employed in the three HT-DFT databases: AFLOW and MP perform
two sequential relaxations, while the OQMD performs sequential relaxations
until the volume change during a relaxation is less than 5\%.

\noindent
\textit{Band gap}:
The distribution of differences in the calculated band gaps is slightly skewed
towards larger band gaps in the OQMD, but this skew is absent in the AFLOW-MP
comparison.
Correspondingly, the median difference in band gaps between AFLOW and MP is
$\sim$0.01~eV, and up to $\sim$0.14~eV for comparisons with OQMD.
The larger band gaps calculated in the OQMD might be due to smaller volumes
from the choice of lower plane wave energy cutoffs.
An increase in the fundamental band gap due to compressive strains (in the
OQMD, due to unresolved Pulay stresses) has been observed in many semiconductor
families~\cite{olsen1978effect, kuo1985effect, wei1999predicted}.
In addition, the spread in the differences in calculated band gaps is quite
large: with an MAD of up to $\sim$0.21~eV and an IQR of up to $\sim$0.36~eV
for comparisons with OQMD.
The spread may be, in addition to the choice of energy cutoff as discussed
above, due to the different ways in which the databases calculate the band gap.
For example, OQMD calculates band gap from the electronic density of states
(DOS), in contrast to AFLOW and MP which calculate it from band dispersions.
The energy grid used for the calculation of DOS and/or $k$-point meshes used
for band structure calculations can also have a notable effect on the precision
and accuracy of the reported band gap.
For instance, while AFLOW and MP both report gaps calculated from band
dispersion calculations, the high-symmetry $k$-path in the Brillouin zone used
for such calculations can be different~\cite{setyawan2010high,
munro2020improved}.

\noindent
\textit{Total magnetization}:
The median differences in AFLOW-MP and MP-OQMD are nearly zero, with reasonably
small MAD values as well.
However, the differences between the magnetization reported in
AFLOW and the other two databases skew towards larger values in AFLOW, with
long tails and correspondingly large dispersions.
The difference between AFLOW and OQMD, in particular, shows an MAD of
$\sim$0.15~$\mu_{\rm B}$/atom and an IQR of $\sim$2.0~$\mu_{\rm B}$/atom.
Further, as noted earlier, a significant fraction of 10--15\% overlapping
records across databases disagree on whether the material has non-zero total
magnetization.
This disagreement may in part be due to different pseudopotential choices for
various elements (and correspondingly different number of valence electrons),
and sampling of different magnetic configurations, the choice of unit cell in
such magnetic configuration sampling, etc.
For instance, AFLOW and MP calculate ferromagnetic configurations for all
materials, and ferrimagnetic and antiferromagnetic configurations for a subset
of materials~\cite{sanvito2017accelerated, horton2019high}, while the OQMD only
calculates ferromagnetic configurations~\cite{kirklin2015open}.
For a given material, since we only compare the lowest-energy configurations
across databases with one another, it is possible that a material is predicted
to be non-magnetic in one database and antiferromagnetic in another database.
Alternately, a ferrimagnetic configuration in one database could be compared to
a ferromagnetic calculation in another, if both converged to finite magnetic
moments.

\subsection{Rank-order comparisons across
properties}\label{ssec:results-rank-comp}

We next seek to make comparisons \textit{across} properties.
Instead of comparing the raw values of the properties directly, we compare
overlapping records using the ordinal rank of the property in each database
being compared (hereafter, referred to as ``percentile rank'').
Comparing the percentile ranks of the properties has a few advantages:
(a) It allows for a single consistent metric for comparison across all four
properties regardless of the magnitude of the actual value and physical units.
(b) It is not affected by many systematic differences, e.g., a constant shift of
0.1~eV in all calculated band gaps in one database.
Such constant shifts in calculated properties do not affect the internal
consistency of a HT-DFT database, and the percentile ranks which are similarly
unaffected capture this property.
(c) It is a robust, uniform, identifier of outliers in calculated properties.

\begin{figure*}[!ht]
\centering
\includegraphics[width=0.9\textwidth]{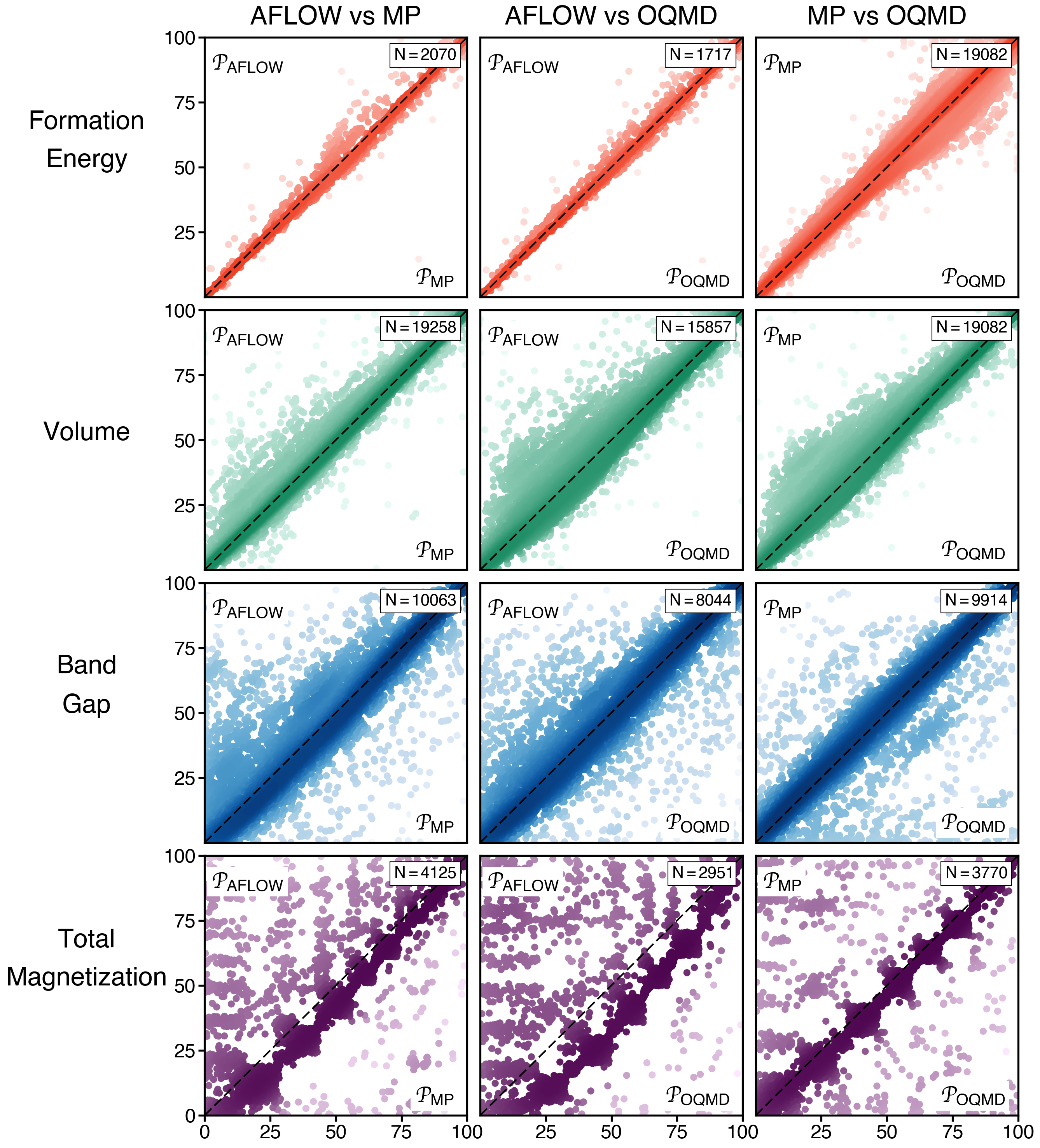}
\caption{Comparison of the calculated properties (formation energy, volume,
  band gap, and total magnetization) over records overlapping across pairwise
  combinations of HT-DFT databases plotted as a percentile rank (i.e., ordinal
  rank of the property in each database being compared).
  A compact line along the diagonal corresponds to perfect correlation between
  the ranked properties.
  Overall, formation energies and volumes show better reproducibility than band
  gaps and magnetizations. The clusters seen in the magnetization comparisons
  correspond to nominally integer values of magnetic
  moments.}\label{fig:pair_percentiles}
\end{figure*}

Figure~\ref{fig:pair_percentiles} consists of percentile rank scatterplots
(closely related to the quantile-quantile or Q-Q plots) of each property of
interest for each database pair.
Note that for band gap (total magnetization), we only include overlapping
records where the two databases being compared both report the material to be
non-metallic (magnetic), to avoid having to rank near-zero or zero values
against one another.
A compact line along the diagonal corresponds to perfect correlation between
the ranked properties, with more diffuse scattering indicating lower levels of
correlation.

\noindent
\textit{Formation energy}:
Of the four properties, formation energy shows the best correlation between
each database pair, consistent with all $r$ and $\rho$ values close to 0.99 in
Table~\ref{tbl:summary_table}.
Nonetheless, there is some off-diagonal scatter for the MP-OQMD comparison for
larger (more positive) values of formation energy that is not found in the
other database pairs.
These calculations correspond to compounds with smaller (positive) formation
energies, where the precision necessary to reliably rank the structure
approaches the accuracy of the calculation.

\noindent
\textit{Volume}:
The percentile rank comparison of volume shows higher off-diagonal scatter than
that seen in comparisons of formation energy.
There is a skew towards higher volumes in AFLOW and MP when compared to OQMD
(scatter towards top-left of the diagonal in the AFLOW-OQMD and MP-OQMD
comparisons), consistent with the discussion around plane wave energy cutoffs
in the previous section.

\noindent
\textit{Band gap}:
The percentile rank comparison of band gap shows even higher off-diagonal
scatter than that observed in comparisons of both formation energy and volume.
In particular, there is meaningful scatter \textit{along the axes},
corresponding to cases where one database predicts the material to have a
near-zero band gap whereas the other database predicts a (much larger) non-zero
band gap.

\noindent
\textit{Total magnetization}:
The percentile rank comparison of total magnetization per formula unit in all
three pairwise comparisons shows a few distinct clusters along the diagonal,
corresponding to nominally integer values of magnetic moment per formula unit.
There is considerable off-diagonal ``bowing'' in the comparisons with AFLOW,
consistent with the distribution of differences between AFLOW and the other two
databases showing a skew towards larger magnetizations in AFLOW \textit{and}
long tails (lower panel in Figure~\ref{fig:pair_histograms}).
In addition, there is considerable off-diagonal scatter (horizontal and
vertical bands in the magnetization panel of
Figure~\ref{fig:pair_percentiles}) indicating significant disagreement between
the values reported in the two databases.

Overall, a comparison of rank-ordered properties across two databases shows
that formation energies and volumes are more easily reproduced than band gaps
and total magnetizations, consistent with correlation coefficients decreasing
from $\sim$0.99 for formation energy to $\sim$0.6 for total magnetization
(Table~\ref{tbl:summary_table}).


\subsection{Reproducibility across materials classes}\label{ssec:results-mat-cls}

Intuitively, we expect the level of agreement among the databases to be a
strong function of materials class.
Therefore, we compare specific subsets of calculations based on various
materials classes to elucidate potential causes of differences.
The materials classes are defined based on chemical composition, the number of
elemental components, the presence of magnetism, band gap, pseudopotential
choices, and space group, as summarized in Table~\ref{tbl:class_definitions}.
For classes defined by the output of a calculation (i.e., those based on
magnetization and band gap), comparisons are only made if both databases agree
that the property has a non-zero value.
Note that according to our definition, the ``Magnetic'' class of materials may
potentially include both ferromagnetic and ferrimagnetic materials, and the
``Non-Magnetic'' class may potentially include both non-magnetic and
antiferromagnetic materials.


\begin{table*}[ht]
\centering
\input{tables/mat_class_defs.tex}
\caption{Definitions for the materials classes used in this
work.}\label{tbl:class_definitions}
\end{table*}


\begin{figure*}[!h]
\centering
\includegraphics[height=0.925\textheight]{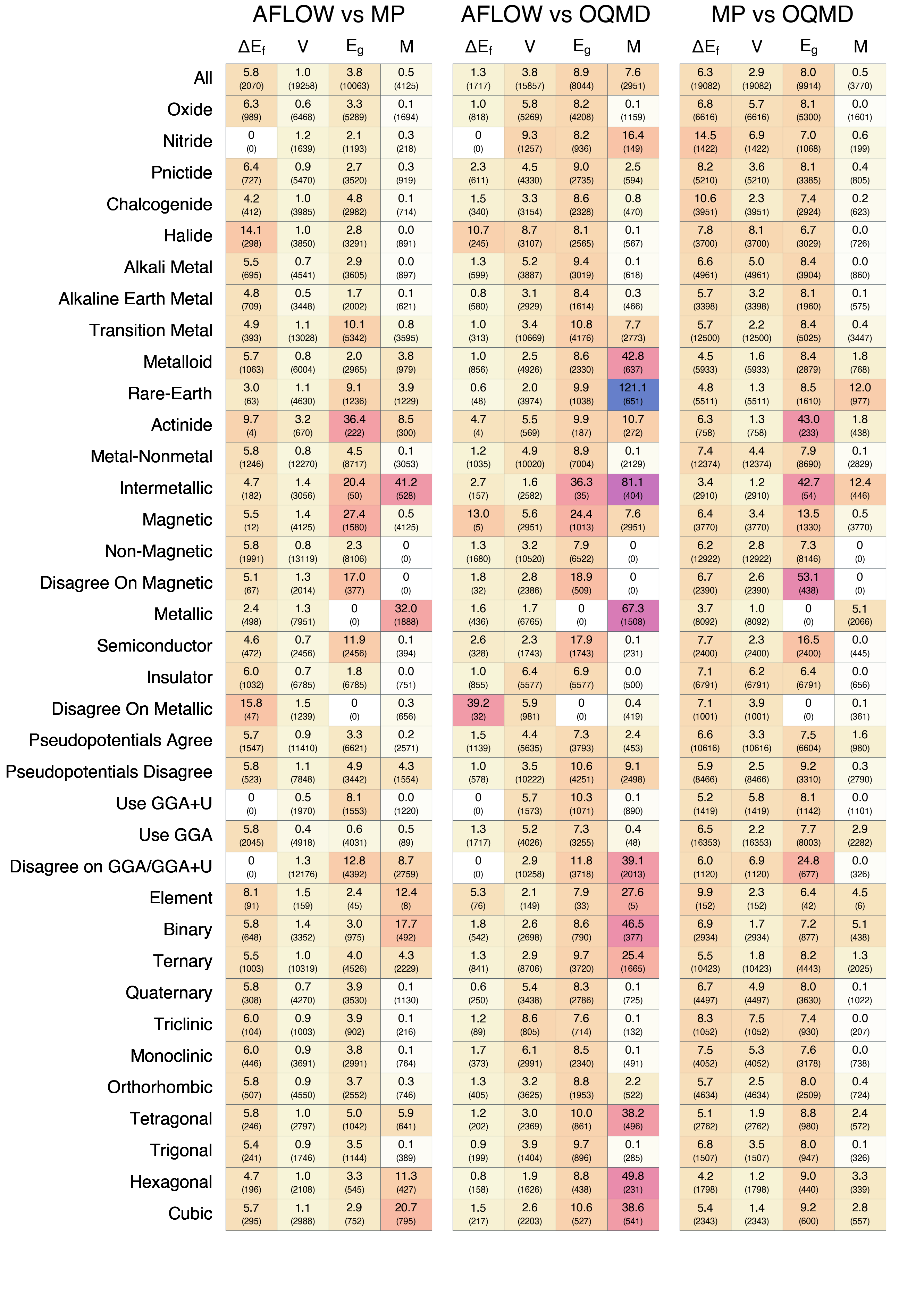}
\caption{Median percent absolute differences between properties (formation
  energy, volume, band gap, total magnetization) calculated in the three
  databases (AFLOW, MP, OQMD), compared two at a time, across various classes
  of materials as defined in Table~\ref{tbl:class_definitions}.
The numbers in parentheses indicate the number of overlapping records belonging
to the respective material class for a given pair of databases.
Trivial comparisons are left blank (e.g., the difference in total magnetization
for non-magnetic compounds).}\label{fig:materials_classes}
\end{figure*}
\clearpage

Figure~\ref{fig:materials_classes} contains the median absolute difference
relative to the mean (MRAD) values for pairwise comparisons between databases,
divided into materials classes as defined in Table~\ref{tbl:class_definitions}.
Cells are colored based on the MRAD value listed.
Empty cells correspond to trivial comparisons (e.g., values of band gap where
both database agree the structure is metallic).
We use MRAD as the metric here to reduce the effect of outliers (as compared to
calculating means) as well as to enable comparisons across properties using the
same metric.
Overall, HT-DFT volumes show the best agreement (lowest MRAD values), from
1--4\%. Band gaps show the worst overall agreement (highest MRAD values),
4--10\% across all pairwise comparisons.
Formation energy comparisons with MP show MRAD values up to 6\%, but the
AFLOW-OQMD MRAD is only 1.3\%.
MRAD values for total magnetization vary highly from 0.5\% for comparisons with
MP to 7.6\% for AFLOW-OQMD.
In all cases, certain materials classes have distinctly higher or lower MRAD
when compared to the MRAD averaged over all materials classes.

\noindent
\textit{Formation Energy}:
In the comparisons with AFLOW, two materials
classes, ``Halides'' and ``Disagree on Metallic'', show the highest MRAD values
of up to 14\% and 40\%, respectively.
The high MRAD in halide formation energies can be understood to result from
post hoc corrections to the effective elemental reference energies performed in
MP and OQMD, but not in AFLOW, for the halide group of elements (see discussion
in Section~\ref{ssec:chem-pot}).
The high MRAD of the ``Disagree on Metallic'' class is likely an artifact
of the small formation energies of the few records ($\sim$30--50) in the
comparison.
As noted earlier, since AFLOW reports notably fewer formation energy values
than the other databases, the comparisons are made with a much smaller set of
records ($\sim$2,000).
Therefore, we ignore here some of the MRAD outliers in cases where the number
of records being compared is very small (e.g., the material class ``Magnetic''
shows an MRAD of 13\% between AFLOW and MP but there are only 5 records in the
comparison).
Further, the formation energies dataset has very few transition metal, rare-earth,
and actinide element-containing compounds (Figures~S3 and S7).
New, different insights are likely to result from a larger dataset.
In the MP-OQMD comparison, with a much larger comparable dataset
($\sim$19,000), the ``Nitride'', ``Pnictide'', and ``Chalcogenide'' material
classes show the highest MRAD values, 14\%, 8\%, and 11\% respectively.
This is partly due to differences in fitted elemental chemical potentials for
pnictogen and chalcogen elements in MP and OQMD (Section~\ref{ssec:chem-pot}).

\noindent
\textit{Volume:}
The best agreement is observed in the AFLOW-MP comparisons, with only the
``Actinide'' material class showing an MRAD greater than 2\%.
For comparisons with OQMD, the MRAD in volume is generally higher---due to the
choice of lower plane wave energy cutoff used for cell relaxation, as discussed
earlier (Section~\ref{ssec:results-diff-dist}).
The highest MRAD values in the comparisons with OQMD volumes are for the
``Nitride'' and ``Halide'' classes ($\sim$7--9\%).
The default plane wave energy cutoffs in the VASP PAW potentials
(\textsf{ENMAX} parameter) for N and F are among the highest (400~eV) of all
elements.
Thus, the lower energy cutoff used by OQMD for relaxation impacts the
calculated volumes of nitrides and fluorides the most (Figures~S8 and S12).
Another material class, ``Triclinic'', shows similarly high MRAD values of
$\sim$8\% in comparisons with OQMD.
Upon examination, we find that most triclinic materials in the comparisons are
oxides, nitrides, and halides, and thus the high MRAD values are due to the
chemical composition of these compounds rather than their crystal symmetry.

\noindent
\textit{Band gap:}
While band gap comparisons show the highest MRAD values across properties, some
materials classes in particular show MRAD values much greater than $\sim$10\%.
Of these, in the ``Intermetallic'' and ``Semiconductor'' material classes, the
MRAD values are expectedly high due to small average band gaps relative to which
differences are reported, even though the absolute differences themselves are
not conspicuously large (Figure~S2).
In other cases, the high MRAD values are a result of
(a) different pseudopotential choices for elements (e.g., Cu/Cu\_pv,
Ce/Ce\_3, Eu/Eu\_2 choices in the ``Disagree on Magnetic'' class for the
MP-OQMD comparison with an MRAD of $\sim$53\%; see Figure~S13),
(b) disagreement on whether to use the GGA or GGA+\textit{U} approach to
calculate properties (e.g., the ``Actinide'' material class with MRAD of up to
43\% in comparisons with MP, the ``Disagree on GGA/GGA+\textit{U}'' class in
all three comparisons with MRAD of 12--25\%), or a combination of both factors
(e.g., for the ``Magnetic'' material class with an MRAD of up to 27\% in
comparisons with AFLOW),
(c) non-overlapping sampling of magnetic configurations across databases.
For instance, the ``Magnetic'' (MRAD of 13--27\% across comparisons) and
``Disagree on Magnetic'' (MRAD of 17--53\% across comparisons) classes may
respectively include comparing ferromagnetic vs ferrimagnetic and non-magnetic
vs antiferromagnetic ground states across two databases (note, however, that
both the ``Magnetic'' and ``Disagree On Magnetic'' comparisons also include
effects from other HT-DFT choices, such as choice of pseudopotential used).
Note also that the errors in band gaps for the ``Use GGA+\textit{U}'' materials
class are larger than those for the ``Use GGA'' materials class across all
three pairwise comparisons, the choice of slightly different effective
\textit{U} values used in the three databases being a likely contributor.
Further discussions of some of the above parameter choices are in
Section~\ref{sec:discussion}.

\noindent
\textit{Total magnetization:}
While MRAD values in the MP-OQMD comparison are generally small ($<5$\%), some
material classes show much higher MRAD values, especially in comparisons with
AFLOW.
As in the case of band gap values, we find these comparisons to be influenced
by pseudopotential choice (of rare-earth elements in particular, e.g., Nd,
Nd\_3, Nd\_3 in AFLOW, MP, and OQMD, respectively; see Figures S10 and S14),
choice of using GGA or GGA+\textit{U} (e.g., MRAD of up to $\sim$40\% in
AFLOW-OQMD comparisons for the ``Disagree on GGA/GGA+\textit{U}'' class), or
both (e.g., the ``Metalloid'' and ``Rare-Earth'' material classes in the
AFLOW-OQMD comparisons, ``Intermetallic'' and ``Metallic'' classes in the
AFLOW-MP and AFLOW-OQMD comparisons).
We note that some other material classes show high MRAD values, e.g.,
``Element'', ``Binary'', ``Ternary'', ``Tetragonal'', ``Hexagonal'', and
``Cubic'' (up to MRAD values up to $\sim$50\%) due to, upon further
examination, the parameter choices discussed above rather than due to number of
components in the compound or crystal symmetry.

Finally, we note that while our scheme of constructing a set of comparable
records across pairs of databases (by matching ICSD IDs exactly) ensures
comparisons between the same initial crystal structures, it excludes a number
of experimentally well-studied materials with multiple ICSD entries associated
with them.
We investigated whether this ``bias away from well-studied materials'' affects
our results by using a larger comparison set constructed by linking very
similar ICSD entries using the crystal structure matching algorithm employed
by the Materials Project (see Section~S-II in the SI).
While some of the quantitative metrics we report varied by a few percent in the
expanded comparison, the overall conclusions remain unchanged (see Tables S-XI,
S-XII, and Figures S15--S18 in the SI), consistent with recent
findings~\cite{matcor-2021}.

%% file: tables/records_tally.tex
\begin{tabular}{lcccccc}
\toprule
{} & AFLOW & MP & OQMD & AFLOW-MP & AFLOW-OQMD & MP-OQMD \\
\midrule
Formation Energy & 2196 & 34907 & 22248 & 2070 & 1717 & 19082 \\
Volume & 21929 & 34907 & 22248 & 19258 & 15857 & 19082 \\
Band Gap & 21921 & 34907 & 22169 & 19253 & 15790 & 19007 \\
Total Magnetization & 21929 & 34907 & 22248 & 19258 & 15857 & 19082 \\
\bottomrule
\end{tabular}

%% file: tables/corr_coeffs.tex
\begin{tabular}{lcccc@{\hskip 12pt}cccc@{\hskip 12pt}cccc}
\toprule
{} & \multicolumn{4}{c}{AFLOW-MP} & \multicolumn{4}{c}{AFLOW-OQMD} & \multicolumn{4}{c}{MP-OQMD} \\
\midrule
{} & MAD & IQR & $r$ & $\rho$ & MAD & IQR & $r$ & $\rho$ & MAD & IQR & $r$ & $\rho$ \\
\midrule
Formation Energy (eV/atom) &  0.105  &  0.173  &  0.99  &  0.99  &  0.019  &  0.036  &  0.99  &  0.99  &  0.087  &  0.168  &  0.99  &  0.99  \\
Volume (\AA$^3$/atom) &  0.180  &  0.389  &  0.98  &  0.99  &  0.647  &  1.117  &  0.97  &  0.97  &  0.512  &  0.902  &  0.98  &  0.98  \\
Band Gap (eV)* &  0.078  &  0.203  &  0.94  &  0.92  &  0.209  &  0.364  &  0.92  &  0.91  &  0.178  &  0.277  &  0.93  &  0.92  \\
Total Magnetization ($\mu_{\rm B}$/f.u.)* &  0.015  &  0.759  &  0.77  &  0.75  &  0.149  &  2.001  &  0.60  &  0.56  &  0.012  &  0.052  &  0.80  &  0.74  \\
\bottomrule
\end{tabular}

%% file: tables/mat_class_defs.tex
\begin{tabular}{@{}ll@{}}
\toprule
\textbf{Class}  & \textbf{Definition} \\
\midrule
Oxide &  Contains O\\
Nitride & Contains N\\
Pnictide & Contains a group 15 element\\ 
Chalcogenide & Contains a group 16 element, except O\\ 
Halide & Contains a group 17 element\\
Alkali Metal & Contains a group 1 element, except H\\
Alkaline Earth Metal & Contains a group 2 element\\
Transition Metal & Contains a $d$-block element\\
Metalloid & Contains B, Si, Ge, As, Sb, or Te\\
Rare-Earth & Contains an element from the lanthanide series\\
Actinide & Contains an element from the actinide series\\
Metal-Nonmetal & Contains at least one metal element \textit{and} at least one of C, N, O, F, P, S, Cl, Se, Br, I\\
Intermetallic & Contains only metallic elements\\
Magnetic & Both databases report a net magnetic moment $>10^{-2}$~$\mu_{\rm B}$/f.u.\\
Non-magnetic & Both databases report no net magnetic moment $>10^{-2}$~$\mu_{\rm B}$/f.u.\\
Disagree on Magnetic & The two databases disagree on whether a net magnetic moment $>10^{-2}$~$\mu_{\rm B}$/f.u. is present\\
Metallic & Both databases predict a band gap of $<10^{-2}$ eV\\
Semiconductor & Both databases predict a band gap between $10^{-2}$ and 1.5~eV\\
Insulator & Both databases predict a band gap larger than 1.5~eV\\
Disagree on Metallic & The two databases disagree on whether a band gap $<10^{-2}$ eV is present\\
Pseudopotentials Agree & Both databases use the same set of pseudopotentials for all elements\\
Pseudopotentials Disagree & The databases use different pseudopotentials for at least one element\\
Use GGA+\textit{U} & Both databases use the GGA+\textit{U} approach\\
Use GGA & Both databases use plain GGA\\
Disagree on GGA/GGA+\textit{U} & One database uses GGA whereas the other uses GGA+\textit{U}\\
Elements & Contains only one element\\
Binaries & Contains two elements\\
Ternaries & Contains three elements\\
Quaternaries & Contains four elements\\
Triclinic & Space group 1--2\\
Monoclinic & Space group 3--15\\
Orthorhombic & Space group 16--74\\
Tetragonal & Space group 75--142\\
Trigonal & Space group 143--167\\
Hexagonal & Space group 168--194\\
Cubic & Space group 195--230\\
\bottomrule
\end{tabular}

%% file: sections/discussion.tex
We discuss some of the most important factors affecting the differences
across HT-DFT calculations of properties below.
Some of the other factors that either have a minor effect (e.g., \textit{post
hoc} calculation of band gap from band dispersions or density of states) or
are specific to a database/property (e.g., plane wave cutoff energy
for full cell relaxations in OQMD) have been discussed in the earlier
sections.

\subsection{Effects of pseudopotential
choice}\label{ssec:discussion-psp-choice}
For nearly all elements, VASP provides multiple PAW potentials to choose
from, with different numbers of electrons in the valence.
The choice of pseudopotential varies across the HT-DFT databases due to
factors such as changes in VASP recommendations and issues of calculation
convergence or reproduction of experimental thermochemical
data~\cite{MPsettings, jain2011high}.
Interestingly, the choice of pseudopotential has minimal effect on the
calculated formation energies and volumes (up to a difference of 1\% in cases
where pseudopotentials do or do not match; see rows ``Pseudopotentials
Agree'' and ``Pseudopotentials Disagree'' in
Figure~\ref{fig:materials_classes}).
On the other hand, the number of valence electrons and consequently the
choice of pseudopotential affects the calculated band gaps and magnetization
values severely.
Especially egregious differences across those properties in material classes
such as ``Rare-Earth'' and ``Magnetic'' (Figure~\ref{fig:materials_classes})
can be directly traced to different pseudopotential choices.
For rare-earth and actinide elements in particular, with $f$-electrons that
are poorly described by DFT~\cite{eyring2002handbook}, using pseudopotentials
that treat $f$-electrons in core or valence can have a significant impact on
the calculated band gap (e.g., ``Intermetallic'' and ``Magnetic'' classes in
Figure~\ref{fig:materials_classes}) and magnetization (e.g., ``Rare-Earth''
and ``Intermetallic'' classes in Figure~\ref{fig:materials_classes}) values.

\subsection{Elemental references and energy corrections}\label{ssec:chem-pot}
The largest disagreements in HT-DFT formation energies can be understood to
result from different elemental reference states and/or post-calculation
energy corrections performed in the databases.
To our knowledge, the formation energies reported in AFLOW use DFT total
energies of the bulk elements as the reference states~\cite{curtarolo2012aflow}.
MP and OQMD both correct DFT-calculated energies to closely reproduce
experimental formation enthalpy data.
While MP adds corrections to the compound formation
energies~\cite{MPsettings, jain2011high}, OQMD fits the elemental reference
energies using a FERE-like approach~\cite{stevanovic2012correcting,
kirklin2015open}.
Such correction schemes involve some more HT-DFT choices:
(a) Should all elemental reference energies and/or compound formation energies
be effectively fit to experimental data or only a subset?
For instance, MP corrects the compound formation energies of nitrides,
fluorides, chlorides, hydrides, sulfides of alkali, alkaline earth, and
aluminum containing compounds~\cite{ong2013python}.
The OQMD fits the reference energies of only elements whose DFT ground states
are poor representation of the experimental reference states (i.e., elements
that are gases or that have a solid-solid phase transition below room
temperature)~\cite{kirklin2015open}.
(b) What experimental thermochemical data should be used such correction
schemes, given a lack of a single, widely-accepted set of standard experimental
dataset for solids?
For instance, MP and OQMD use experimental formation energies from different
sources to fit elemental reference energies: MP uses data from Materials
Thermochemistry~\cite{kubaschewski1993}, while OQMD uses data from SGTE
SUBstance Database (SSUB)~\cite{sgte1999} in addition to others (see
Refs.~\onlinecite{jain2011high, kirklin2015open} for details of the fitting
data used in the two databases).
Some other standard reference databases are also widely used, such as the
NIST-JANAF Thermochemical Tables~\cite{chase1998nist}.
Since a given material may have experimental data in one or more such
reference databases of experimental properties, the choice of the source of
experimental data affects the fitted formation energies in HT-DFT databases,
even in cases where other parameters such as pseudopotentials used are held
constant.
This effect of fitted elemental reference states is shown in the calculated
formation energies averaged over compounds containing each element in
Figures~S3, S7, and S11.

\subsection{GGA vs.\ GGA+U approach}\label{ssec:gga+u}
One of the ways to treat the issue of over-delocalization in DFT is to use the
DFT+\textit{U} approach~\cite{anisimov1991, kulik2015perspective} (or
``GGA+\textit{U}'' when used with GGA).
Similar to the case of fitting elemental references, using the GGA+\textit{U}
approach requires additional HT-DFT choices.
(a) Whether or not to use GGA+\textit{U} for calculating properties of a given
material.
All three HT-DFT databases have slightly different sets of compounds for which
the GGA+\textit{U} approach is applied.
The OQMD uses GGA+\textit{U} only for oxides of certain 3$d$ transition metals
(the V--Cu series) and actinide metals~\cite{kirklin2015open}.
MP uses GGA+\textit{U} for oxides, fluorides, and sulfides of a larger set of
transition metals, but not actinides~\cite{jain2011high}.
AFLOW applies it to an even larger set of compounds, nearly all those
containing $d$- or $f$-block elements~\cite{Calderon2015233}.
(b) What effective \textit{U} value should be used for each element?
The three HT-DFT databases all use different effective \textit{U} values for
each element, obtained either from previous work (OQMD) or in-house
parameterization by fitting to experimental data (AFLOW and
MP)~\cite{setyawan2010high, jain2011formation}.
Such choices around when to use the GGA+\textit{U} approach to calculate a
compound and what effective \textit{U} value to use can impact some properties
more than others, e.g., discrepancies in total magnetization values in the
AFLOW-OQMD comparisons, particularly for ``Rare-Earth'', ``Intermetallic'', and
``Metallic'' classes.
For some properties, such as formation energies, \textit{post hoc} corrections
are required to maintain consistency between those calculated using the GGA and
GGA+\textit{U} approaches, especially while constructing phase diagrams
involving compounds calculated using the two different approaches.
Such corrections are obtained by fitting to experimental reaction energies, and
can be different between HT-DFT databases based on the source of such reaction
energies.



%% file: sections/conclusion.tex
Recent years have seen a dramatic increase in the application of
informatics methods for materials development, using high-throughput DFT
data.
Several prominent HT-DFT databases exist and each uses different input
parameters and post-processing techniques to calculate materials properties.
Quantifying the uncertainty in calculated properties due to such parameter
choices is therefore crucial to understanding the reproducibility and
interoperability of such data.
In this work, we centralize data from three of the largest HT-DFT databases,
AFLOW, Materials Project, and OQMD, into a common data repository, allowing
records to be accurately compared.
We then compare four properties---formation energy, volume, band gap,
and total magnetization---of materials calculated in each of the HT-DFT
databases using the same initial crystal structure.

Our comparisons show that formation energy and volume are more easily
reproduced than band gap and total magnetization.
Interestingly, we find that the average difference in calculated
properties across two HT-DFT databases is comparable to that between DFT and
experiment: up to 0.105~eV/atom for formation energy, 4\% for volume,
0.21~eV for band gap, and 0.15 $\mu_{\rm B}$/formula unit for total
magnetization.
Further, certain input parameter choices disproportionately affect HT-DFT
properties of particular classes of materials, e.g. choice of planewave cutoff
on formation energies and volumes of oxides and halides, and the choice of
pseudopotential on the band gaps and magnetization of rare-earth compounds.
Our results inform users of the variability to account for in reported
materials properties, especially when using data from multiple HT-DFT
databases in their own analyses.
In addition, our quantitative uncertainty estimates can directly aid materials
informatics efforts, e.g., for separation of model uncertainty and inherent
noise in data.

As HT-DFT databases continue to mature, systematic comparisons,
interoperability, and standardization of calculations become increasingly
crucial.
Efforts to improve the interoperability of materials databases, e.g., by the
development of a common data schema by the OPTiMaDe
consortium~\cite{optimade2020}, are already ongoing.
Toward improving the standardization of calculations, HT-DFT choices and
reproducibility in particular, we list a few recommendations for
next-generation and new iterations of current HT-DFT databases:
\begin{enumerate}
  \item [(a)] \textit{In-depth, versioned documentation} of the various parameter
    choices made in a high-throughput project, including the data-driven
    rationale for the choices, if any.
  \item [(b)] \textit{Visibility for possible uncertainty} in reported
    properties (in both the web and programmatic interfaces used to interact
    with HT-DFT data) for which HT-DFT choices are expected to have a
    significant impact.
    Further, we recommend providing estimated uncertainties in calculated
    properties, either determined from literature references (e.g., this work),
    or from in-house investigations (e.g., by performing a set of HT-DFT
    calculations with different input parameters as part of a sensitivity
    analysis).
  \item [(c)] \textit{Community-led initiative to reach a consensus} on which
    HT-DFT choices ought to be standardized (e.g., energy cutoffs, fitting sets
    for empirical corrections, post-processing steps to determine properties
    such as band gap) and which HT-DFT choices could be a source of greater
    scientific insight if they were more diverse (e.g., DFT codes,
    pseudopotentials, DFT exchange-correlation functionals).
\end{enumerate}


%% file: sections/metric_defs.tex
The definitions of statistical quantities and their symbols used in this work
throughout are as follows ($x_i$ and $y_i$ refer to the two sets of data being
compared, e.g.\ from two different databases):
\begin{enumerate}
  \item Median difference ($\widetilde{\Delta x}$):
    \begin{equation}\label{eqn:md}
      \widetilde{\Delta x} = {\rm median}(x_i - y_i)
    \end{equation}
  \item Median absolute difference (MAD):
    \begin{equation}\label{eqn:mad}
      {\rm MAD} = {\rm median}\big(\left|x_i - y_i\right|\big)
    \end{equation}
  \item Interquartile range (IQR):
    \begin{equation}\label{eqn:iqr}
      {\rm IQR} = {\rm Q}_3 - {\rm Q}_1
    \end{equation}
    where Q$_1$ and Q$_3$ are the first and third quartiles (25th and 75th
    percentiles), respectively.
  \item Median relative absolute difference (MRAD):
    \begin{equation}\label{eqn:mpad}
      {\rm MRAD} = {\rm median}\left(\frac{|x_i - y_i|}{|x_i + y_i|/2}
      \times 100\right)
    \end{equation}
  \item Pearson correlation coefficient ($r$):
    \begin{equation}\label{eqn:pr}
      r(x, y) = \frac{\sum_i^n (x_i - \bar{x})(y_i - \bar{y})}{
      \sqrt{\sum_i^n (x_i - \bar{x})^2} \sqrt{\sum_i^n (y_i - \bar{y})^2}}
    \end{equation}
    where $\bar{x} = \frac{1}{n}\sum_i^n x_i$ is the sample mean, and $n$ is
    the sample size.
  \item Spearman's rank correlation coefficient ($\rho$) is defined as the
    Pearson correlation coefficient between rank variables $x^{\rm R}_i$ and
    $y^{\rm R}_i$ corresponding to raw data values $x_i$ and $y_i$,
    respectively:
    \begin{equation}\label{eqn:rho}
      \rho (x, y)  = r(x^{\rm R}, y^{\rm R})
    \end{equation}
\end{enumerate}

%% file: sections/htdft_tables.tex
Given the popularity of DFT as a method generating materials data, a number of
HT-DFT databases with various focus areas have emerged; a list of exemplars is
given in Table~\ref{tbl:dbs}.
Similarly, codes for the management of these databases, including the worfklows
to generate the data, have been developed by several groups around the world,
such as the packages listed in Table~\ref{tbl:codes}.

\begin{table*}[htp]
\caption{A selection of publicly-available HT-DFT databases.}\label{tbl:dbs}
\input{tables/htdft_dbs}
\end{table*}

\begin{table*}[htp]
\caption{A selection of software tools for automating HT-DFT workflows and
property calculations.}\label{tbl:codes}
\input{tables/htdft_software}

\end{table*}

%% file: tables/htdft_dbs.tex
\begin{tabular}{l l p{0.2\textwidth} p{0.30\textwidth} c}
\toprule
Database & Link & Materials & Properties & Reference \\
\midrule
Aflowlib & \href{http://materials.duke.edu/aflow.html}{materials.duke.edu/aflow.html} & inorganic solids & electronic structure, thermodynamics & [\onlinecite{curtarolo2012aflowlib}] \\
Alloy database & \href{http://alloy.phys.cmu.edu}{alloy.phys.cmu.edu} & intermetallics & structure, cohesive energies & [\onlinecite{widom2005stability}] \\
CatApp & \href{https://slac.stanford.edu/~strabo/catapp}{slac.stanford.edu/$\sim$strabo/catapp} & molecules on surfaces & reaction/activation energies & [\onlinecite{hummelshoj2012catapp}] \\
CCCBDB & \href{https://cccbdb.nist.gov}{cccbdb.nist.gov} & atoms, molecules & thermochemical properties & [\onlinecite{cccbdb}] \\
CMR & \href{https://cmr.fysik.dtu.dk}{cmr.fysik.dtu.dk} & perovskites, 2D materials & energetics, electronic structure & [\onlinecite{landis2012computational}] \\
CompES-X & \href{https://compes-x.nims.go.jp/index_en.html}{compes-x.nims.go.jp} & inorganic solids & electronic structure & \\
Crystalium & \href{http://crystalium.materialsvirtuallab.org/}{crystalium.materialsvirtuallab.org} & elemental solids & surface, grain boundary energetics & [\onlinecite{tran2016surface}] \\
CEP & & organic photovoltaics & HOMO-LUMO energies & [\onlinecite{hachmann2011harvard}] \\
JARVIS-DFT & \href{https://jarvis.nist.gov}{jarvis.nist.gov} & 2D/solid inorganics & elastic, thermoelectric properties & [\onlinecite{choudhary2020joint}] \\
Materials Cloud & \href{https://www.materialscloud.org/discover}{www.materialscloud.org} & porous, 2D materials & structural, topological  properties & [\onlinecite{talirz2020materials}] \\
Materials Project & \href{https://materialsproject.org}{materialsproject.org} & inorganic solids & mechanical, dielectric, piezoelectric & [\onlinecite{jain2013commentary}] \\
NoMaD & \href{https://nomad-coe.eu/}{nomad-coe.eu} & inorganic solids & raw DFT calculation files & \\
NRELMatDB & \href{https://materials.nrel.gov/}{materials.nrel.gov} & inorganic solids & quasiparticle energies & [\onlinecite{stevanovic2012correcting}] \\
OQMD & \href{http://oqmd.org}{oqmd.org} & inorganic solids & energetics, electronic structure & [\onlinecite{saal2013materials}] \\
phonondb & \href{http://phonondb.mtl.kyoto-u.ac.jp}{phonondb.mtl.kyoto-u.ac.jp} & inorganic solids & phonons, thermal properties & \\
TE Design Lab & & semiconductors & electronic, thermoelectric properties & [\onlinecite{gorai2016te}] \\
\bottomrule
\end{tabular}

%% file: tables/htdft_software.tex
\begin{tabular}{@{}lllll@{}}
\toprule
Code & Functionality & Link & Reference \\
\midrule
\multicolumn{4}{c}{High-throughput workflows} \\
 AFLOW & calculation setup, submission & \href{http://materials.duke.edu/aflow.html}{materials.duke.edu/aflow.html} & [\onlinecite{setyawan2010high}] \\
 AiiDA & calculation setup, submission, storage & \href{http://aiida.net}{aiida.net} & [\onlinecite{pizzi2016aiida}] \\
 ASE & calculation setup, submission, analysis & \href{https://wiki.fysik.dtu.dk/ase}{wiki.fysik.dtu.dk/ase} & [\onlinecite{larsen2017atomic}] \\
 MPInterfaces & surface calculation setup, analysis & \href{https://github.com/henniggroup/MPInterfaces}{github.com/henniggroup/MPInterfaces} & [\onlinecite{mathew2016mpinterfaces}] \\
 pymatgen & calculation setup, analysis & \href{https://pymatgen.org/}{pymatgen.org} & [\onlinecite{ong2013python}] \\
 qmpy & calculation setup, management, analysis & \href{https://pypi.org/project/qmpy/}{pypi.org/project/qmpy} & [\onlinecite{kirklin2015open}] \\
\midrule
\multicolumn{4}{c}{Calculation of properties/model building} \\[0.5\baselineskip]
Amp & atomistic potentials & \href{https://amp.readthedocs.io/en/latest/}{amp.readthedocs.io} & [\onlinecite{khorshidi2016amp}] \\
ATAT & cluster expansions & \href{https://www.brown.edu/Departments/Engineering/Labs/avdw/atat/}{brown.edu/Departments/Engineering/Labs/avdw/atat} & [\onlinecite{van2002alloy}] \\
atomate & electronic structure, dielectric tensors & \href{https://atomate.org/}{atomate.org} & [\onlinecite{mathew2017atomate}] \\
CALYPSO & crystal structure prediction & \href{http://www.calypso.cn/}{www.calypso.cn} & [\onlinecite{wang2012calypso}] \\
MAST & defects, diffusion & \href{https://pythonhosted.org/MAST/}{pythonhosted.org/MAST} & [\onlinecite{mayeshiba2017materials}] \\
Mint & crystal structure utilities & \href{https://github.com/materials/mint}{github.com/materials/mint} & \\
phonopy & phonons, high-temperature properties & \href{https://atztogo.github.io/phonopy/}{atztogo.github.io/phonopy} & [\onlinecite{togo2015first}] \\
SeeK-path & high-symmetry paths in Brillouin zone & \href{https://www.materialscloud.org/work/tools/seekpath}{materialscloud.org/work/tools/seekpath} & [\onlinecite{hinuma2017band}] \\
SLUSCHI & melting temperatures & \href{https://blogs.brown.edu/qhong/?page_id=102}{blogs.brown.edu/qhong/?page\_id=102} & [\onlinecite{hong2016user}] \\
USPEX & crystal structure prediction& \href{https://uspex-team.org/en/uspex/overview}{uspex-team.org/en/uspex/overview} & [\onlinecite{glass2006uspex}] \\
Xtalopt & crystal structure prediction  & \href{http://xtalopt.github.io/}{xtalopt.github.io} & [\onlinecite{lonie2011xtalopt}] \\
\bottomrule
\end{tabular}

%% file: sections/data_management.tex
As described in the main text, here we focus on three high-throughput DFT
databases: Automatic FLOW (AFLOW), the Materials Project (MP) and the Open
Quantum Materials Database (OQMD).
MP and OQMD distribute the HT-DFT data and related content via the Creative
Commons Attribution 4.0 License
(\url{https://creativecommons.org/licenses/by/4.0}).
Individual AFLOW records contain the following license statement:
``The data included in the AFLOW repository is free for scientific, academic,
and non-commercial purposes, and any other use is prohibited.''

The primary challenge in comparing calculations from different HT-DFT databases
is to collect the relevant database entries and determine equivalency between
unique calculations of the same structure.
This section outlines the workflow of ingesting data from AFLOW, Materials
Project, and OQMD into a central repository, all post-extraction processing of
the data, the method for curating the post-processed data, and the method of
determining whether two records were comparable.

\subsection{Querying Data from HT-DFT Databases}
To create a complete and searchable system of records, entries from each HT-DFT
database were imported into the a central repository, after standardization into
the Physical Information File (PIF) format~\cite{michel2016beyond} using the
\texttt{pypif}~\cite{pypif} package (see Section~\ref{sec:example_pifs} for a
few example PIFs with data queried from HT-DFT databases). The data in this work
represents the data found on the public-facing HT-DFT databases, aggregated in
June 2021.\\

Each HT-DFT database was queried independently using their respective APIs:
\begin{enumerate}
  \item AFLOW: The query was performed using the AFLUX
    API~\cite{curtarolo2012aflowlib}, retrieving all records in the
    \texttt{icsd} catalog.
    The query resulted in 60,324 records, which were stored as PIF objects.
  \item Materials Project: The query was performed using the MPRester
    API~\cite{ong2015materials}, retrieving all records with non-empty
    \texttt{icsd\_ids} field.
    The Materials Project (MP) release log
    (\url{https://matsci.org/t/materials-project-database-release-log})
    indicates that the version corresponding to when data in this work was
    queried is v2019.11.
    The query resulted in 48,833 records, which were stored as PIF objects.
    In addition, some normalized properties such as per-atom volume and
    per-formula-unit magnetization that were not queried directly but
    calculated from queried data were added to the PIF objects.
  \item OQMD: The query was performed using qmpy and a version 1.2 of the
    underlying database, filtering on \texttt{FormationEnergy} objects
    associated with converged "\texttt{static}" \texttt{Calculation} objects
    and \texttt{Entry} objects with "\texttt{icsd}" in the "\texttt{path}"
    field.
    The query resulted in 37,989 records, which were stored as PIF objects.
    In addition, some properties that were not directly queried but retrieved as
    metadata, such as labels of PAW potentials and crystal system, were added to
    the PIF objects.
\end{enumerate}

\subsection{Aggregation of HT-DFT Datasets}
The HT-DFT database-specific tags for each of the queried properties are given
in Table~\ref{tbl:prop_name_table}.
All three datasets are aggregated in the PIF format, using a unique key for each
property irrespective of the field names queried from the HT-DFT databases.
For instance, the number of atoms in the unit cell is stored as
\texttt{number\_of\_atoms} for all three databases despite varying labels (e.g.,
\texttt{nsites}, \texttt{natoms}, \texttt{calculation\_\_output\_\_natoms})
across databases.
The aggregated dataset is available in the \texttt{data} folder in this GitHub
repository: \url{https://github.com/CitrineInformatics-ERD-public/htdft-uq}.

\begin{table*}[htp]
\caption{HT-DFT database-specific tags for properties compared in this work.
  In cases where they were not directly queried, normalized values such as
  per-atom values for ``volume'' and per-formula unit values for ``total
  magnetization'' were calculated from the non-normalized queried
data.}\label{tbl:prop_name_table}
\input{tables/property_tags}
\end{table*}

\subsection{Structure Equivalency}\label{sec:aliasing}

In order to perform a fair comparison between two HT-DFT databases, it is
essential to generate a set of equivalent records for each pair of databases
considered.
We use the ICSD Collection Code(s) (hereafter, ``ICSD ID(s)'') in the metadata
of each entry to generate a set of comparable records.

\subsubsection{Exact ICSD ID Matching}\label{sssec:exact_icsd_id}

The main text contains analysis only for records with the exact same ICSD ID
across the two databases being compared at a time, ensuring that the crystal
structure of the materials being compared are the same.
Since the ICSD ID of each material is retrieved as part of the metadata queried
from each of the three HT-DFT databases, materials can be ``matched'' across
databases by directly comparing their ICSD IDs.

\subsubsection{Aliasing for Multiple ICSD IDs Per
Record}\label{sssec:multi_icsd_id}

Since the former process of exact ICSD ID matching results in a smaller set of
records (less than 50\%) when compared to the total ICSD entries in each
database, we investigated if our results hold on a larger comparison set
generated by linking similar ICSD entries.
We describe the process of linking similar ICSD entries below and provide
tables and figures corresponding to this larger dataset in
Section~\ref{sec:multi-icsd-id-results}.

For the process of linking similar ICSD entries, we use the structure
comparison and matching algorithm implemented within the Materials Project.
The process is involves the following two steps:\\

\noindent
\textbf{STEP 1. Annotation of extracted data with ``ICSD UID''}

\begin{enumerate}
  \item \textit{Generation of a set of ICSD UIDs from Materials Project:}
    An ICSD UID is defined as a set of ICSD Collection Codes (ICSD IDs)
    belonging to the same material.
    Materials Project already groups ICSD IDs per material but due to the
    affine mapping based structure matching implemented (\textit{J. Appl.
    Cryst.} \textbf{39}, 6--16 (2006) DOI:
    \href{https://doi.org/10.1107/S0021889805032450}{10.1107/S0021889805032450}),
    the material can be part of two different ``groups of equivalent
    structures''.
    For example, there are three different records for AgO in Materials Project
    that share the ICSD ID 60625.
    This task in the curation pipeline thus aggregates the ICSD IDs from all
    three such records to generate a super ``ICSD UID''.
    For instance, GaN (mp-830) was associated with 10 ICSD IDs, and the
    corresponding ICSD UID is the set of all 10 ICSD IDs:
    ``187047--190412--67781--156260--41546--157511--248504--191770--185155--290614''.
  \item \textit{Annotation of extracted records with an ICSD UID:}
    First, every record retrieved from Materials Project is matched to an ICSD
    UID from the set above, depending on the ICSD IDs in the record (the set of
    ICSD IDs in the record will be a subset of exactly one ICSD UID).
    Second, it is attempted to match every record retrieved from AFLOW and OQMD
    to ICSD UID. If no match is found, the ICSD ID in such a record constitutes
    a new ICSD UID.
\end{enumerate}

The resulting extracted properties dictionary has the following format:

\begin{verbatim}
{
     AFLOW: [
       {
          db_id: 1,
          prop_name_1: prop_val_1,
          prop_name_2: prop_val_2,
          icsd_uid: icsd_id_1-icsd_id_6-icsd_id_7,
          ...
       },
       {
          db_id: 2,
          icsd_uid: icsd_id_2-icsd_id_3,
          ...
       },
       {
          db_id: 3,
          icsd_uid: icsd_id_1-icsd_id_6-icsd_id_7,
          ...
       },
       ...
       ...
     ],

     MP: [
       ...
       ...
     ],
  ...
  ...
}
\end{verbatim}

\noindent
\textbf{STEP 2. Reverse-mapping properties to ICSD UIDs}\\

This step involves ``inverting'' the dictionary of extracted properties above
such that the ICSD UIDs are the keys and a list of data records corresponding
to each UID is the value.
The inverted dictionary at this step has the following format:

\begin{verbatim}
{
    AFLOW: {
      icsd_id_1-icsd_id_6-icsd_id_7: [
        {
           db_id: 1,
           prop_name_1: prop_val_1,
           prop_name_2: prop_val_2,
        },
        {
           db_id: 3,
           ...
        }
      ],
      icsd_id_2-icsd_id_3: [
        {
           db_id: 2,
           ...
        }
      ],
    },

    MP: {
      ...
      ...
    },
  ...
  ...
}
\end{verbatim}

The rest of this document, except Section~\ref{sec:multi-icsd-id-results},
reports data curation performed on datasets generated via ``Exact ICSD ID
Matching'' approach.
For the analysis on the larger datasets generated by linking similar ICSD IDs
using the structure matching method within MP (presented in
Section~\ref{sec:multi-icsd-id-results}), curation steps as described below
were used as well.

\subsection{Data Curation} 

\subsubsection{Removing composition inconsistencies}

From the ICSD UID to properties dictionary above, an ICSD UID key is removed
if the entries within it do not have matching compositions.
This process is done first within each of the three databases (discarded
entries in Table~\ref{tbl:composition-disagreement}), and then for UIDs common
to pair-wise combinations of the databases (discarded entries in
Table~\ref{tbl:comp-disagree-across-dbs}).
Most records filtered out at this step are materials with different number of H
and Li atoms (e.g., \ce{BaGaH4} vs \ce{BaGaH5}) or small changes in composition
(e.g., \ce{Y3Fe29} vs \ce{Y3Fe31}).

\subsubsection{Filtering for the lowest energy entry per ICSD UID}

For each ICSD UID, since there may exist multiple entries (calculations) in
every database, only the entry with the lowest
\texttt{total\_energy\_per\_atom} value is retained from this step onward.

\subsubsection{Removing records with unphysical properties}

At this step, any records with unphysical values of certain properties are
removed.
This includes all boride formation energies from AFLOW, due to an error in the
B chemical potential (this bug was discovered in the course of this work and
confirmed by the AFLOW developers~\cite{aflowlib_priv_comm}).
Beyond AFLOW borides, unphysical properties are defined as
per-atom formation energies outside $-5$ to $+5$~eV/atom, per-atom volumes
above 150~{\AA}$^3$/atom, for all three databases (entries discarded at this
step in Table~\ref{tbl:physically-unreasonable}).

\subsubsection{Converting magnetizations into absolute values}

Finally, all \texttt{total\_magnetization\_per\_atom} values in all three
databases are converted into absolute values.

When querying the Materials Project for total magnetization via the RESTful
API, the value returned is not the total magnetic moment of the unit cell, as
documented~\cite{MPrestful_doc}, but rather the per formula unit value.
This issue has been communicated to the Materials Project development team.
The magnetization values queried from Materials Project were normalized
suitably.\\

Lastly, we identify the largest outliers for each property in each pairwise
database comparison \textit{post-curation}: formation energy in
Table~\ref{tbl:outliers_formation_energy}, volume in
Table~\ref{tbl:outliers_volume}, band gap in Table~\ref{tbl:outliers_band_gap},
and total magnetization in Table~\ref{tbl:outliers_magnetization}.\\

\begin{table*}[htp]
\caption{Instances where composition did not match for records \textit{within a
database}.}\label{tbl:composition-disagreement}
\input{tables/within_db_comp_mm}
\end{table*}

\clearpage
\begin{longtable*}{l l c}
\caption{Instances where composition did not match for records \textit{across
two databases being compared}.}\label{tbl:comp-disagree-across-dbs}
\input{tables/across_db_comp_mm}
\end{longtable*}

\begin{table*}[htp]
\caption{Records with physically unreasonable values of formation energy
  (outside the [$-5$, $+5$] eV/atom window) and/or volume
($>$150~\AA$^3$/atom).}\label{tbl:physically-unreasonable}
\input{tables/unphys_props}
\end{table*}

\begin{table*}
\caption{Top ten outliers in calculated formation energy across pairwise
comparisons of databases.}\label{tbl:outliers_formation_energy}
\input{tables/outliers__formation_energy_per_atom}
\end{table*}

\begin{table*}
\caption{Top ten outliers in calculated volume across pairwise comparisons of
databases.}\label{tbl:outliers_volume}
\input{tables/outliers__volume_per_atom}
\end{table*}

\begin{table*}
\caption{Top ten outliers in calculated band gap across pairwise comparisons of
databases.}\label{tbl:outliers_band_gap}
\input{tables/outliers__band_gap}
\end{table*}

\begin{table*}
\caption{Top ten outliers in calculated total magnetization across pairwise
comparisons of databases.}\label{tbl:outliers_magnetization}
\input{tables/outliers__total_magnetization_per_atom}
\end{table*}

%% file: tables/property_tags.tex
\begin{tabular}{ l c c c }
\toprule
Property & AFLOW & Materials Project & OQMD \\
\midrule
Record ID & \texttt{auid} & \texttt{material\_id} & \texttt{entry\_id} \\
Record URL & \texttt{aurl} & \texttt{-} & \texttt{-} \\
Composition & \texttt{compound} & \texttt{unit\_cell\_formula} & \texttt{composition\_\_formula} \\
Number of atoms & \texttt{natoms} & \texttt{nsites} & \texttt{output\_\_natoms} \\
Total energy (/unit cell) & \texttt{energy\_cell} & \texttt{final\_energy} & \texttt{calculation\_\_energy} \\
Total energy (/atom) & \texttt{energy\_atom} & \texttt{final\_energy\_per\_atom} & \texttt{calculation\_\_energy\_pa} \\
Formation energy & \texttt{enthalpy\_formation\_atom} & \texttt{formation\_energy\_per\_atom} & \texttt{formation\_\_delta\_e} \\
Convex hull distance & \texttt{-} & \texttt{e\_above\_hull} & \texttt{formation\_\_stability} \\
Volume (/unit cell) & \texttt{volume\_cell} & \texttt{volume} & \texttt{output\_\_volume} \\
Volume (/atom) & \texttt{volume\_atom} & \texttt{-} & \texttt{output\_\_volume\_pa} \\
Magnetization (/unit cell) & \texttt{spin\_cell} & \texttt{total\_magnetization} & \texttt{-} \\
Magnetization (/atom) & \texttt{spin\_atom} & \texttt{-} & \texttt{calculation\_\_magmom\_pa} \\
Band gap & \texttt{Egap} & \texttt{band\_gap} & \texttt{calculation\_\_band\_gap} \\
Space group ITC \# & \texttt{spacegroup\_relax} & \texttt{spacegroup} & \texttt{spacegroup\_\_number} \\
Crystal system & \texttt{lattice\_system\_relax} & \texttt{crystal\_system} & \texttt{-} \\
Pseudopotentials & \texttt{species\_pp} & \texttt{pseudo\_potential} & \texttt{calculation\_\_settings} \\
\bottomrule
\end{tabular}

%% file: tables/within_db_comp_mm.tex
\begin{tabular}{l l c}
\toprule
Database & Compositions & ICSD UID \\
\midrule
AFLOW & \ce{Ba2CeCl7}, \ce{AgSbTe2}, \ce{YIO}, \ce{Bi}, \ce{Ba2LaCl7}, \ce{Ba2YCl7}, \ce{YBrO} & 0 \\
AFLOW & \ce{Cr2Te4O11}, \ce{LiB} & 1 \\
AFLOW & \ce{SrGaH5}, \ce{SrGaH4} & 240697 \\
AFLOW & \ce{BaGaH4}, \ce{BaGaH5} & 240693 \\
\midrule
OQMD & \ce{ZnCoCuAg}, \ce{LiMgSnPd} & 16478 \\
\bottomrule
\end{tabular}

%% file: tables/across_db_comp_mm.tex
\\
\toprule
Databases & Compositions & ICSD UID \\
\midrule
\endfirsthead

\toprule
Databases & Compositions & ICSD UID \\
\midrule
\endhead

\multicolumn{3}{c}{Continued on the next page\ldots} \\
\bottomrule
\endfoot

\bottomrule
\endfoot

AFLOW-MP & \ce{LaHO2}, \ce{LaO2} & 60675 \\
AFLOW-MP & \ce{XeSb2F10}, \ce{Xe2Sb4F19} & 157664 \\
AFLOW-MP & \ce{Ba5P3HO13}, \ce{Ba5P3O13} & 62283 \\
AFLOW-MP & \ce{Li2ScP2HO8}, \ce{Li2Sc(PO4)2} & 409955 \\
AFLOW-MP & \ce{KAl2P2H5O11}, \ce{KAl2P2H4O11} & 407355 \\
AFLOW-MP & \ce{RbHO}, \ce{Rb2O2} & 61048 \\
AFLOW-MP & \ce{Ho2B3HO8}, \ce{Ho2B3O8} & 413928 \\
AFLOW-MP & \ce{Rb3H(SO4)2}, \ce{Rb3(SO4)2} & 60050 \\
AFLOW-MP & \ce{Ta3Al4HO14}, \ce{Ta3Al4O14} & 67673 \\
AFLOW-MP & \ce{Gd2CBr2}, \ce{GdBr} & 47226 \\
AFLOW-MP & \ce{HO3I}, \ce{O3I} & 26621 \\
AFLOW-MP & \ce{LiZnBi}, \ce{ZnBi} & 100115 \\
AFLOW-MP & \ce{Y3Fe31}, \ce{Y3Fe29} & 107259 \\
AFLOW-MP & \ce{RbOs2HO9}, \ce{RbOs2O9} & 20611 \\
AFLOW-MP & \ce{Ba5Re3O17}, \ce{Ba5Re3O16} & 100777 \\
AFLOW-MP & \ce{Na3VP2HO9}, \ce{Na3VP2O9} & 50760 \\
AFLOW-MP & \ce{Na3TiHF8}, \ce{Na3TiF8} & 14131 \\
AFLOW-MP & \ce{C2NHS2(O2F3)2}, \ce{C2NS2(O2F3)2} & 50524 \\
AFLOW-MP & \ce{BaHF3}, \ce{BaF3} & 35409 \\
AFLOW-MP & \ce{HgHO4Cl}, \ce{HgO4Cl} & 29038 \\
AFLOW-MP & \ce{LiTa3(Bi2O7)2}, \ce{Ta3(Bi2O7)2} & 415141 \\
AFLOW-MP & \ce{YHO2}, \ce{YO2} & 28442 \\
AFLOW-MP & \ce{Dy2B3HO8}, \ce{Dy2B3O8} & 413927 \\
AFLOW-MP & \ce{CaHOCl}, \ce{CaOCl} & 24403 \\
AFLOW-MP & \ce{NdAl2}, \ce{Nd2Al} & 608745 \\
AFLOW-MP & \ce{Ga(Bi4O7)3}, \ce{Ga(Bi3O5)4} & 68648 \\
AFLOW-MP & \ce{HoHO2}, \ce{HoO2} & 2944 \\
AFLOW-MP & \ce{Y12(ReC3)5}, \ce{Y12Re5C6} & 658805 \\
AFLOW-MP & \ce{SnPHO3}, \ce{SnPO3} & 25034 \\
AFLOW-MP & \ce{LiCH3O3}, \ce{LiCH2O3} & 109604 \\
AFLOW-MP & \ce{CaCuAsHO5}, \ce{CaCuAsO5} & 64694 \\
AFLOW-MP & \ce{K2Cr2AsHO10}, \ce{K2Cr2AsO10} & 30533 \\
AFLOW-MP & \ce{SrHF3}, \ce{SrF3} & 35408 \\
AFLOW-MP & \ce{NdMoHO15I4}, \ce{NdMoO15I4} & 281173 \\
AFLOW-MP & \ce{BaAl5HO9}, \ce{BaAl5O9} & 33282 \\
AFLOW-MP & \ce{SrNH}, \ce{SrN} & 410656 \\
AFLOW-MP & \ce{Si2N3H}, \ce{Si2N3} & 202970 \\
AFLOW-MP & \ce{La3TaH(O2Cl)3}, \ce{La3Ta(O2Cl)3} & 62189 \\
AFLOW-MP & \ce{NaCHO2}, \ce{NaCO2} & 109643 \\
AFLOW-MP & \ce{LaNHO4}, \ce{LaNO4} & 413563 \\
AFLOW-MP & \ce{Ba5Cr3HO13}, \ce{Ba5Cr3O13} & 21034 \\
AFLOW-MP & \ce{H2O}, \ce{H7O4} & 27844 \\
AFLOW-MP & \ce{CdHOCl}, \ce{CdOCl} & 26752 \\
AFLOW-MP & \ce{Tl3HS2O9}, \ce{Tl3S2O9} & 35358 \\
AFLOW-MP & \ce{CdNHO4}, \ce{CdNO4} & 35355 \\
AFLOW-MP & \ce{MnP2HO7}, \ce{MnP2O7} & 415152 \\
\midrule
AFLOW-OQMD & \ce{LiB}, \ce{MnCoSiO4} & 2 \\
AFLOW-OQMD & \ce{Y3Fe31}, \ce{Y3Fe29} & 107259 \\
AFLOW-OQMD & \ce{Ga(Bi4O7)3}, \ce{Ga(Bi3O5)4} & 68648 \\
AFLOW-OQMD & \ce{H2O}, \ce{H7O4} & 27844 \\
\midrule
MP-OQMD & \ce{Bi3(PO5)2}, \ce{SmPd} & 107679 \\
MP-OQMD & \ce{SrGaH5}, \ce{SrGaH4} & 240697 \\
MP-OQMD & \ce{BaGaH5}, \ce{BaGaH4} & 240693 \\

%% file: tables/unphys_props.tex
\begin{tabular}{l l r c}
\toprule
Database & Composition & Property value & ICSD UID \\
\midrule
\multicolumn{4}{c}{Formation Energy (eV/atom)} \\
AFLOW & \ce{SiO2} & $-11.950$ & 170547 \\
MP & \ce{Ta} & $5.113$ & 54207 \\
OQMD & \ce{CuO2} & $1126.321$ & 54126 \\
OQMD & \ce{SiO2} & $84.972$ & 155252 \\
OQMD & \ce{Sc5Ga3} & $8.575$ & 165189 \\
\midrule
\multicolumn{4}{c}{Volume (\AA$^3$/atom)} \\
MP & \ce{HoVO4} & $364.707$ & 152694 \\
MP & \ce{HgH3IO6} & $188.228$ & 409499 \\
MP & \ce{Rb} & $523.351$ & 109016 \\
MP & \ce{CaC2} & $896.388$ & 252718 \\
MP & \ce{EuAg} & $962.227$ & 58257 \\
MP & \ce{SnSe} & $182.422$ & 52425 \\
MP & \ce{Na} & $349.842$ & 70067 \\
MP & \ce{TcB} & $733.428$ & 168896 \\
MP & \ce{K} & $433.990$ & 157565 \\
MP & \ce{TlCl2} & $386.012$ & 20762 \\
MP & \ce{Ta} & $309.658$ & 54207 \\
MP & \ce{Fe3C} & $217.282$ & 76827 \\
MP & \ce{Cu} & $603.475$ & 150682 \\
\bottomrule
\end{tabular}

%% file: tables/outliers__formation_energy_per_atom.tex
\begin{tabular}{l l r r r c}
\toprule
Databases & Composition & $\Delta E_{\rm f}^1$ & $\Delta E_{\rm f}^2$ & $\Delta^{1-2}$ & ICSD UID \\
 & & \multicolumn{3}{c}{(eV/atom)} & \\
\midrule
AFLOW-MP & \ce{AlPO4} & $-2.780$ & $-0.307$ & $-2.474$ & 162670 \\
AFLOW-MP & \ce{SiO2} & $-1.041$ & $-2.870$ & $1.829$ & 25632 \\
AFLOW-MP & \ce{O} & $0.004$ & $1.669$ & $-1.665$ & 92775 \\
AFLOW-MP & \ce{AlClO} & $-1.741$ & $-2.789$ & $1.048$ & 27812 \\
AFLOW-MP & \ce{Bi} & $0.135$ & $1.090$ & $-0.954$ & 51675 \\
AFLOW-MP & \ce{YAl3} & $0.515$ & $-0.437$ & $0.952$ & 58220 \\
AFLOW-MP & \ce{PCl5} & $-0.192$ & $-1.061$ & $0.869$ & 76731 \\
AFLOW-MP & \ce{K2BeO2} & $-1.259$ & $-2.124$ & $0.865$ & 23633 \\
AFLOW-MP & \ce{Li4P2O7} & $-1.899$ & $-2.762$ & $0.863$ & 39814 \\
AFLOW-MP & \ce{RbC8} & $0.751$ & $-0.030$ & $0.781$ & 200563 \\
\midrule
AFLOW-OQMD & \ce{AlPO4} & $-2.780$ & $-0.237$ & $-2.543$ & 162670 \\
AFLOW-OQMD & \ce{O} & $0.004$ & $1.395$ & $-1.391$ & 92775 \\
AFLOW-OQMD & \ce{MgO} & $-2.877$ & $-1.879$ & $-0.998$ & 181459 \\
AFLOW-OQMD & \ce{AlClO} & $-1.741$ & $-2.608$ & $0.867$ & 27812 \\
AFLOW-OQMD & \ce{RbC8} & $0.751$ & $-0.040$ & $0.791$ & 200563 \\
AFLOW-OQMD & \ce{K2BeO2} & $-1.259$ & $-2.002$ & $0.743$ & 23633 \\
AFLOW-OQMD & \ce{SrPSe3} & $-0.277$ & $-1.016$ & $0.739$ & 412766 \\
AFLOW-OQMD & \ce{Li4P2O7} & $-1.899$ & $-2.634$ & $0.735$ & 39814 \\
AFLOW-OQMD & \ce{PCl5} & $-0.192$ & $-0.908$ & $0.716$ & 76731 \\
AFLOW-OQMD & \ce{Mg3P2O8} & $-2.346$ & $-2.930$ & $0.584$ & 9849 \\
\midrule
MP-OQMD & \ce{MnNiAs} & $3.547$ & $-0.249$ & $3.796$ & 161716 \\
MP-OQMD & \ce{Nd3PbN} & $-0.945$ & $1.319$ & $-2.264$ & 76397 \\
MP-OQMD & \ce{IrC4} & $0.981$ & $3.161$ & $-2.180$ & 181498 \\
MP-OQMD & \ce{Nd3SnN} & $-1.029$ & $1.112$ & $-2.141$ & 76398 \\
MP-OQMD & \ce{Sm3AlN} & $-0.894$ & $1.109$ & $-2.003$ & 52640 \\
MP-OQMD & \ce{LiNO3} & $0.690$ & $-1.252$ & $1.942$ & 33661 \\
MP-OQMD & \ce{Pr3AlN} & $-0.788$ & $1.101$ & $-1.888$ & 52639 \\
MP-OQMD & \ce{Cr3O} & $1.971$ & $0.141$ & $1.830$ & 15904 \\
MP-OQMD & \ce{Eu(Cu2Sn)2} & $1.296$ & $-0.322$ & $1.617$ & 416796 \\
MP-OQMD & \ce{GdMg2Ag} & $1.502$ & $-0.102$ & $1.605$ & 107733 \\
\bottomrule
\end{tabular}

%% file: tables/outliers__volume_per_atom.tex
\begin{tabular}{l l r r r c}
\toprule
Databases & Composition & $V^1$ & $V^2$ & $\Delta^{1-2}$ & ICSD UID \\
 & & \multicolumn{3}{c}{(\AA$^3$/atom)} & \\
\midrule
AFLOW-MP & \ce{Bi} & $31.51$ & $103.72$ & $-72.21$ & 51675 \\
AFLOW-MP & \ce{MnNiAs} & $15.70$ & $84.29$ & $-68.59$ & 161716 \\
AFLOW-MP & \ce{Hg} & $28.86$ & $95.21$ & $-66.35$ & 79804 \\
AFLOW-MP & \ce{HoS} & $39.85$ & $103.31$ & $-63.46$ & 66357 \\
AFLOW-MP & \ce{SrPSe3} & $89.79$ & $27.94$ & $61.85$ & 412766 \\
AFLOW-MP & \ce{NbTeBr3} & $75.01$ & $34.02$ & $40.99$ & 35376 \\
AFLOW-MP & \ce{CeBr3} & $67.72$ & $31.52$ & $36.20$ & 31582 \\
AFLOW-MP & \ce{Se} & $63.04$ & $31.90$ & $31.14$ & 150731 \\
AFLOW-MP & \ce{FeSeBr7} & $66.74$ & $37.16$ & $29.58$ & 39528 \\
AFLOW-MP & \ce{SnSe} & $27.09$ & $56.42$ & $-29.34$ & 71338 \\
\midrule
AFLOW-OQMD & \ce{SrPSe3} & $89.79$ & $27.65$ & $62.14$ & 412766 \\
AFLOW-OQMD & \ce{NbTeBr3} & $75.01$ & $30.71$ & $44.30$ & 35376 \\
AFLOW-OQMD & \ce{CeBr3} & $67.72$ & $31.01$ & $36.71$ & 31582 \\
AFLOW-OQMD & \ce{H} & $37.20$ & $2.73$ & $34.47$ & 28465 \\
AFLOW-OQMD & \ce{CsTl} & $14.22$ & $47.71$ & $-33.49$ & 165344 \\
AFLOW-OQMD & \ce{Se} & $63.04$ & $31.39$ & $31.65$ & 150731 \\
AFLOW-OQMD & \ce{FeSeBr7} & $66.74$ & $37.22$ & $29.52$ & 39528 \\
AFLOW-OQMD & \ce{CaPSe3} & $53.48$ & $25.50$ & $27.98$ & 412765 \\
AFLOW-OQMD & \ce{SnSe} & $27.09$ & $52.60$ & $-25.52$ & 71338 \\
AFLOW-OQMD & \ce{TiO2} & $35.12$ & $11.39$ & $23.73$ & 97008 \\
\midrule
MP-OQMD & \ce{Bi} & $103.72$ & $31.53$ & $72.20$ & 51675 \\
MP-OQMD & \ce{MnNiAs} & $84.29$ & $14.74$ & $69.55$ & 161716 \\
MP-OQMD & \ce{Hg} & $95.21$ & $27.10$ & $68.11$ & 79804 \\
MP-OQMD & \ce{CoO2} & $57.55$ & $11.71$ & $45.84$ & 89837 \\
MP-OQMD & \ce{H2} & $43.46$ & $5.38$ & $38.07$ & 28344 \\
MP-OQMD & \ce{Cd6Sb5} & $62.94$ & $31.30$ & $31.65$ & 52832 \\
MP-OQMD & \ce{Cl2} & $70.98$ & $41.47$ & $29.50$ & 22406 \\
MP-OQMD & \ce{LiBH4} & $37.31$ & $9.87$ & $27.44$ & 168803 \\
MP-OQMD & \ce{HfPd5} & $41.53$ & $15.69$ & $25.84$ & 168289 \\
MP-OQMD & \ce{Xe} & $83.51$ & $58.86$ & $24.65$ & 9786 \\
\bottomrule
\end{tabular}

%% file: tables/outliers__band_gap.tex
\begin{tabular}{l l r r r c}
\toprule
Databases & Composition & $E_{\rm g}^1$ & $E_{\rm g}^2$ & $\Delta^{1-2}$ & ICSD UID \\
 & & \multicolumn{3}{c}{(eV)} & \\
\midrule
AFLOW-MP & \ce{CeF3} & $5.87$ & $0.00$ & $5.87$ & 42470 \\
AFLOW-MP & \ce{LiAlPHO5} & $0.02$ & $5.72$ & $-5.70$ & 68921 \\
AFLOW-MP & \ce{Na2PHO3} & $0.00$ & $5.58$ & $-5.58$ & 155976 \\
AFLOW-MP & \ce{KCeF4} & $5.44$ & $0.00$ & $5.44$ & 23229 \\
AFLOW-MP & \ce{RbYbF3} & $1.09$ & $6.52$ & $-5.43$ & 49590 \\
AFLOW-MP & \ce{K5NaCe2S6O24} & $5.49$ & $0.08$ & $5.41$ & 281576 \\
AFLOW-MP & \ce{CsYbF3} & $1.78$ & $7.05$ & $-5.26$ & 49579 \\
AFLOW-MP & \ce{BaTm2F8} & $1.97$ & $7.24$ & $-5.26$ & 20103 \\
AFLOW-MP & \ce{CePO4} & $5.23$ & $0.00$ & $5.23$ & 184550 \\
AFLOW-MP & \ce{Mg3P2O8} & $0.00$ & $5.18$ & $-5.18$ & 9849 \\
\midrule
AFLOW-OQMD & \ce{BeF2} & $8.04$ & $0.00$ & $8.04$ & 173557 \\
AFLOW-OQMD & \ce{KYb3F10} & $0.97$ & $8.44$ & $-7.47$ & 28258 \\
AFLOW-OQMD & \ce{NaCaAlF6} & $7.12$ & $0.00$ & $7.12$ & 80542 \\
AFLOW-OQMD & \ce{H24OsC8N2F6} & $0.00$ & $6.68$ & $-6.68$ & 151185 \\
AFLOW-OQMD & \ce{YbCl3O12} & $0.00$ & $6.38$ & $-6.38$ & 85762 \\
AFLOW-OQMD & \ce{LiAlPHO5} & $0.02$ & $6.20$ & $-6.18$ & 68921 \\
AFLOW-OQMD & \ce{LiEuP4O12} & $0.33$ & $6.41$ & $-6.08$ & 416878 \\
AFLOW-OQMD & \ce{Na2PHO3} & $0.00$ & $6.07$ & $-6.07$ & 155976 \\
AFLOW-OQMD & \ce{CoSiH12O6F6} & $0.00$ & $5.96$ & $-5.96$ & 2900 \\
AFLOW-OQMD & \ce{CsYbF3} & $1.78$ & $7.73$ & $-5.94$ & 49579 \\
\midrule
MP-OQMD & \ce{KYb3F10} & $0.00$ & $8.44$ & $-8.44$ & 28258 \\
MP-OQMD & \ce{BeF2} & $7.96$ & $0.00$ & $7.96$ & 173557 \\
MP-OQMD & \ce{KCeF4} & $0.00$ & $7.75$ & $-7.75$ & 23229 \\
MP-OQMD & \ce{H2} & $0.00$ & $7.22$ & $-7.22$ & 28539 \\
MP-OQMD & \ce{NaCaAlF6} & $7.11$ & $0.00$ & $7.11$ & 80542 \\
MP-OQMD & \ce{EuMgF4} & $0.29$ & $7.22$ & $-6.94$ & 86246 \\
MP-OQMD & \ce{CsEuF3} & $0.00$ & $6.93$ & $-6.93$ & 49577 \\
MP-OQMD & \ce{H24OsC8(NF3)2} & $0.17$ & $6.68$ & $-6.51$ & 151185 \\
MP-OQMD & \ce{Yb(ClO4)3} & $0.00$ & $6.46$ & $-6.46$ & 85763 \\
MP-OQMD & \ce{LiEu(PO3)4} & $0.00$ & $6.41$ & $-6.41$ & 416878 \\
\bottomrule
\end{tabular}

%% file: tables/outliers__total_magnetization_per_atom.tex
\begin{tabular}{l l r r r c}
\toprule
Databases & Composition & $M^1$ & $M^2$ & $\Delta^{1-2}$ & ICSD UID \\
 & & \multicolumn{3}{c}{($\mu_{\rm B}$/formula unit)} & \\
\midrule
AFLOW-MP & \ce{YbMn28} & $124.89$ & $7.16$ & $117.72$ & 643923 \\
AFLOW-MP & \ce{Pr6Mn23} & $121.85$ & $32.42$ & $89.43$ & 643337 \\
AFLOW-MP & \ce{BaMn28} & $87.51$ & $8.61$ & $78.90$ & 615966 \\
AFLOW-MP & \ce{Yb6Mn23} & $105.23$ & $29.18$ & $76.05$ & 643920 \\
AFLOW-MP & \ce{Ba6Co25S27} & $66.99$ & $0.00$ & $66.99$ & 71939 \\
AFLOW-MP & \ce{Mn20W3C6} & $74.10$ & $16.64$ & $57.46$ & 618279 \\
AFLOW-MP & \ce{Mn20Mo3C6} & $73.37$ & $16.53$ & $56.84$ & 618260 \\
AFLOW-MP & \ce{Gd8Rh5C12} & $56.95$ & $0.15$ & $56.80$ & 617956 \\
AFLOW-MP & \ce{Nd12Co6Sn} & $56.33$ & $0.00$ & $56.33$ & 240094 \\
AFLOW-MP & \ce{Yb6Co30P19} & $56.51$ & $1.12$ & $55.39$ & 67950 \\
\midrule
AFLOW-OQMD & \ce{EuMn28} & $125.94$ & $4.13$ & $121.82$ & 631390 \\
AFLOW-OQMD & \ce{Gd13Ge6O31F} & $90.97$ & $0.01$ & $90.96$ & 62329 \\
AFLOW-OQMD & \ce{Pr6Mn23} & $121.85$ & $33.03$ & $88.82$ & 643337 \\
AFLOW-OQMD & \ce{BaMn28} & $87.51$ & $5.58$ & $81.93$ & 615966 \\
AFLOW-OQMD & \ce{Yb6Mn23} & $105.23$ & $30.78$ & $74.45$ & 643920 \\
AFLOW-OQMD & \ce{Ba6Co25S27} & $66.99$ & $2.23$ & $64.76$ & 71939 \\
AFLOW-OQMD & \ce{Nd12Co6Sn} & $56.33$ & $0.01$ & $56.32$ & 240094 \\
AFLOW-OQMD & \ce{Gd7Pd3} & $54.76$ & $0.89$ & $53.86$ & 104112 \\
AFLOW-OQMD & \ce{Gd7CoI12} & $51.02$ & $0.00$ & $51.01$ & 245279 \\
AFLOW-OQMD & \ce{ThMn12} & $52.10$ & $2.25$ & $49.86$ & 104986 \\
\midrule
MP-OQMD & \ce{Gd13Ge6O31F} & $91.00$ & $0.01$ & $90.99$ & 62329 \\
MP-OQMD & \ce{ZnFe16Ni7O32} & $38.00$ & $94.00$ & $-56.00$ & 182238 \\
MP-OQMD & \ce{Eu7Au3} & $50.72$ & $0.01$ & $50.71$ & 611842 \\
MP-OQMD & \ce{Gd6Zn23} & $42.88$ & $0.01$ & $42.87$ & 636504 \\
MP-OQMD & \ce{Gd6C3Cl5} & $42.69$ & $0.00$ & $42.68$ & 202547 \\
MP-OQMD & \ce{Gd10S19} & $42.00$ & $0.00$ & $42.00$ & 416804 \\
MP-OQMD & \ce{Ba8Eu7Cl34} & $45.00$ & $4.02$ & $40.98$ & 408479 \\
MP-OQMD & \ce{Mn9Au31} & $0.28$ & $39.05$ & $-38.77$ & 58552 \\
MP-OQMD & \ce{Eu5Pd2} & $36.02$ & $0.00$ & $36.02$ & 631525 \\
MP-OQMD & \ce{Eu5Pt2} & $35.94$ & $0.00$ & $35.94$ & 631557 \\
\bottomrule
\end{tabular}

%% file: sections/dhf_bimodal.tex
\begin{figure}[ht]
\includegraphics[height=0.35\textheight]{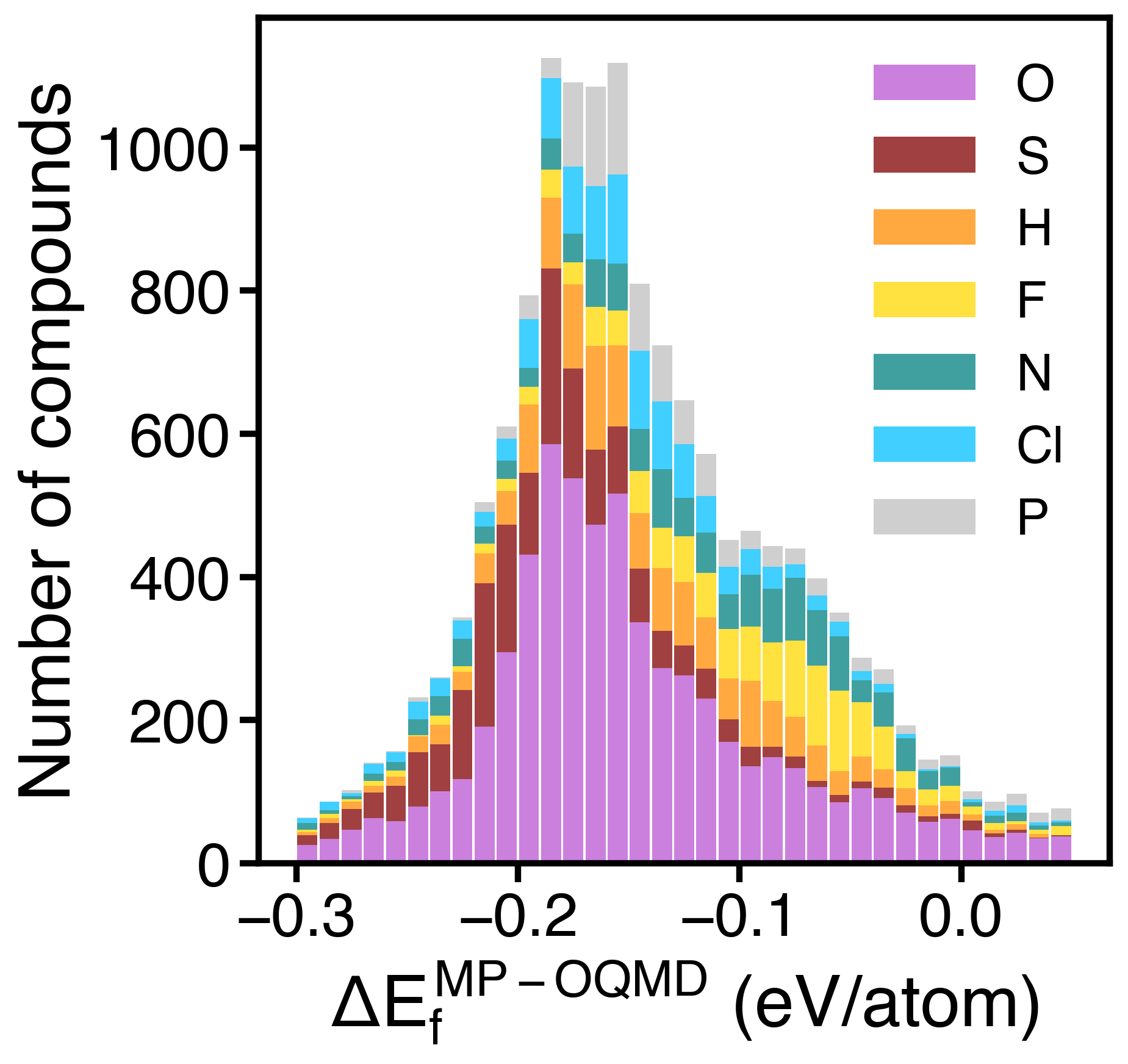}
\caption{Distribution of compounds in the MP-OQMD comparison set, near the peak
  at $\Delta E_{\rm f} = 0.2$~eV/atom in the histogram of formation energy
  differences. Only compounds containing the top few most-frequently-occurring
elements are shown.}\label{fig:dhf_bimodal}
\end{figure}

%% file: sections/pmc_mad.tex
\begin{figure}[ht]
\includegraphics[height=0.75\textheight]{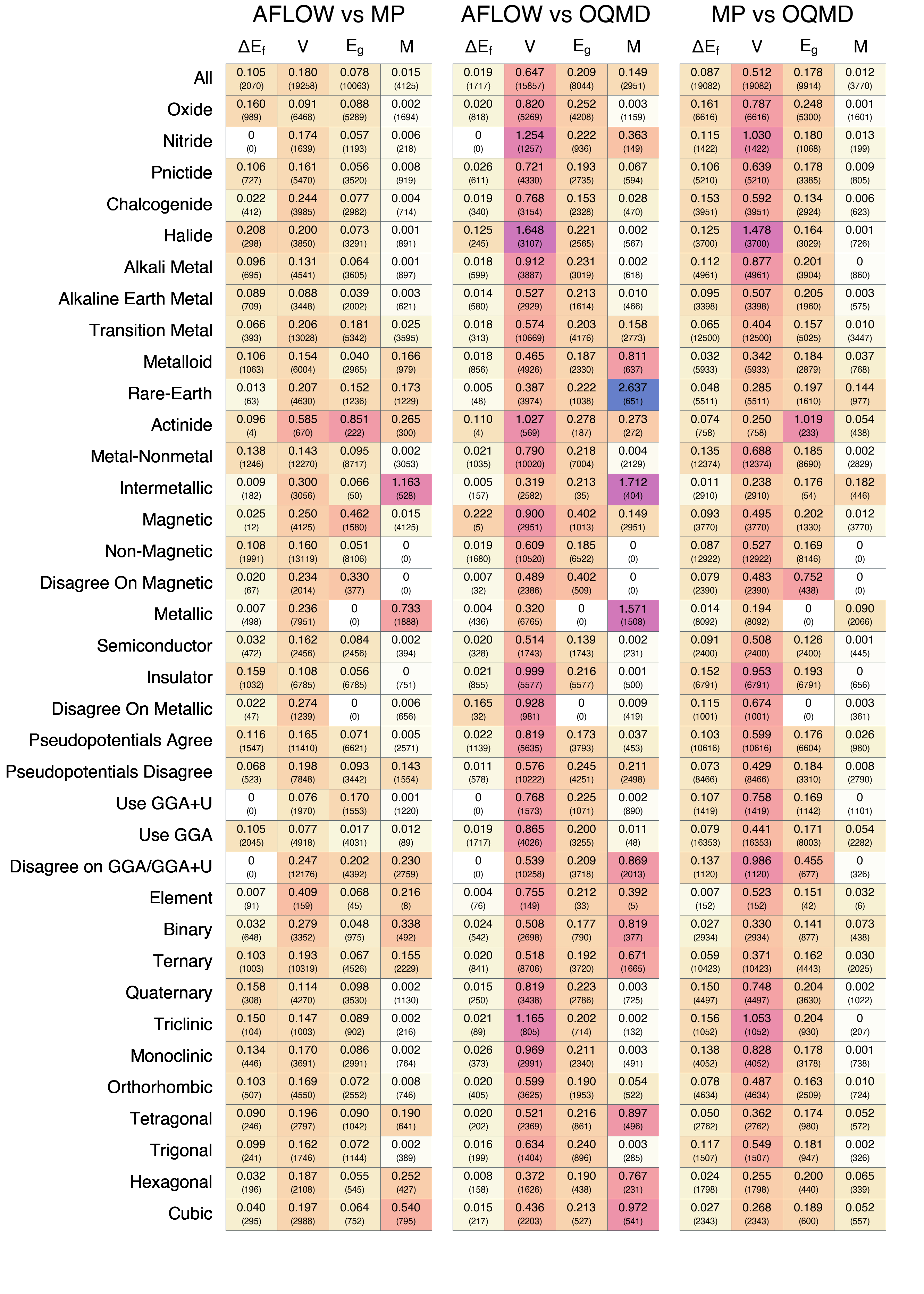}
\caption{Median absolute differences between properties (formation energy,
  volume, band gap, total magnetization are in units of eV/atom,
  {\AA}$^3$/atom, eV, and $\mu_{\rm B}$/formula unit, respectively) calculated
  in the three databases (AFLOW, MP, OQMD), compared pairwise, across various
  classes of materials as defined in Table~III of the main text.
  The numbers in parentheses indicate the number of overlapping records
  belonging to the respective material class for a given pair of databases.
  Trivial comparisons are left blank (e.g., the difference in total
magnetization for non-magnetic compounds).}\label{fig:mat-cls-mad}
\end{figure}

%% file: sections/element_wise.tex
\begin{figure}[htb]
  \includegraphics[width=1.00\linewidth]{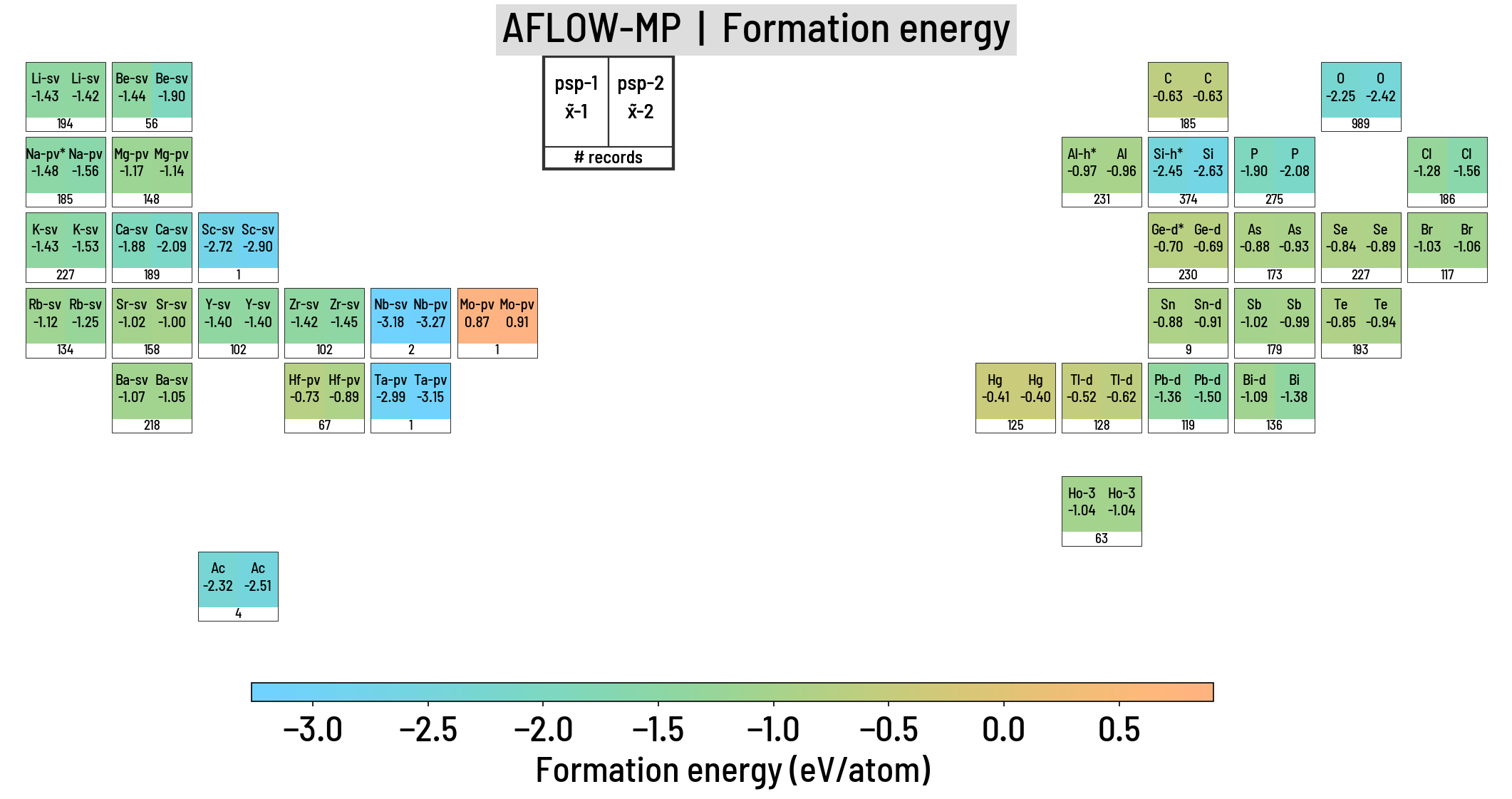}
  \caption{Median values of formation energy for compounds containing a certain
    element in the periodic table, for a comparison of AFLOW and MP.
    The VASP PAW potential used for each element and the number of records in
    each comparison are indicated (* indicates more than one pseudopotential
    used in the database overall for that element).}\label{fig:AFLOW-MP-dhf}
\end{figure}
\begin{figure}[htb]
  \includegraphics[width=1.00\linewidth]{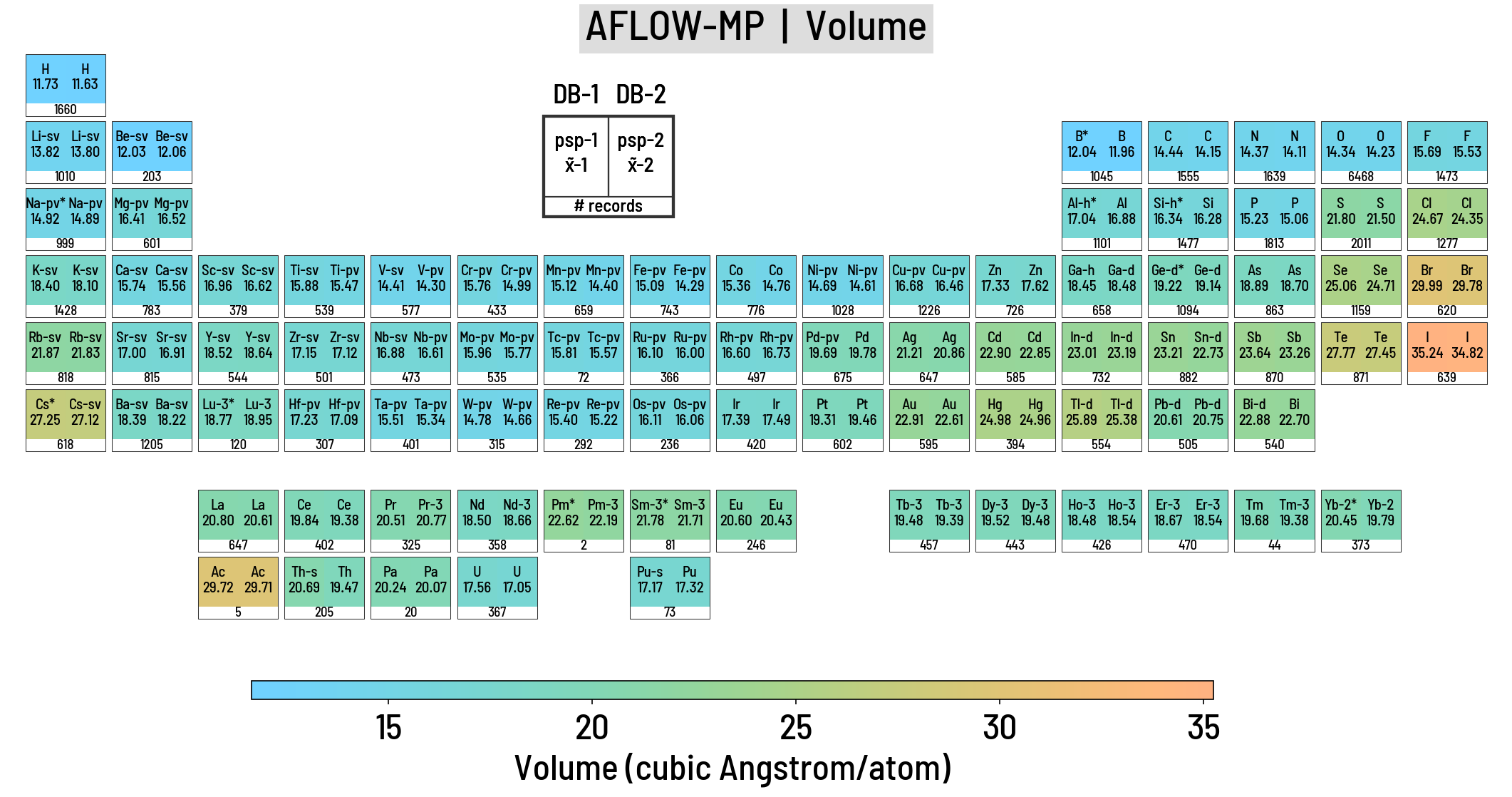}
  \caption{Median values of per-atom volume for compounds containing a certain
    element in the periodic table, for a comparison of AFLOW and MP.
    The VASP PAW potential used for each element and the number of records in
    each comparison are indicated (* indicates more than one pseudopotential
    used in the database overall for that element).}\label{fig:AFLOW-MP-vol}
\end{figure}
\begin{figure}[htb]
  \includegraphics[width=1.00\linewidth]{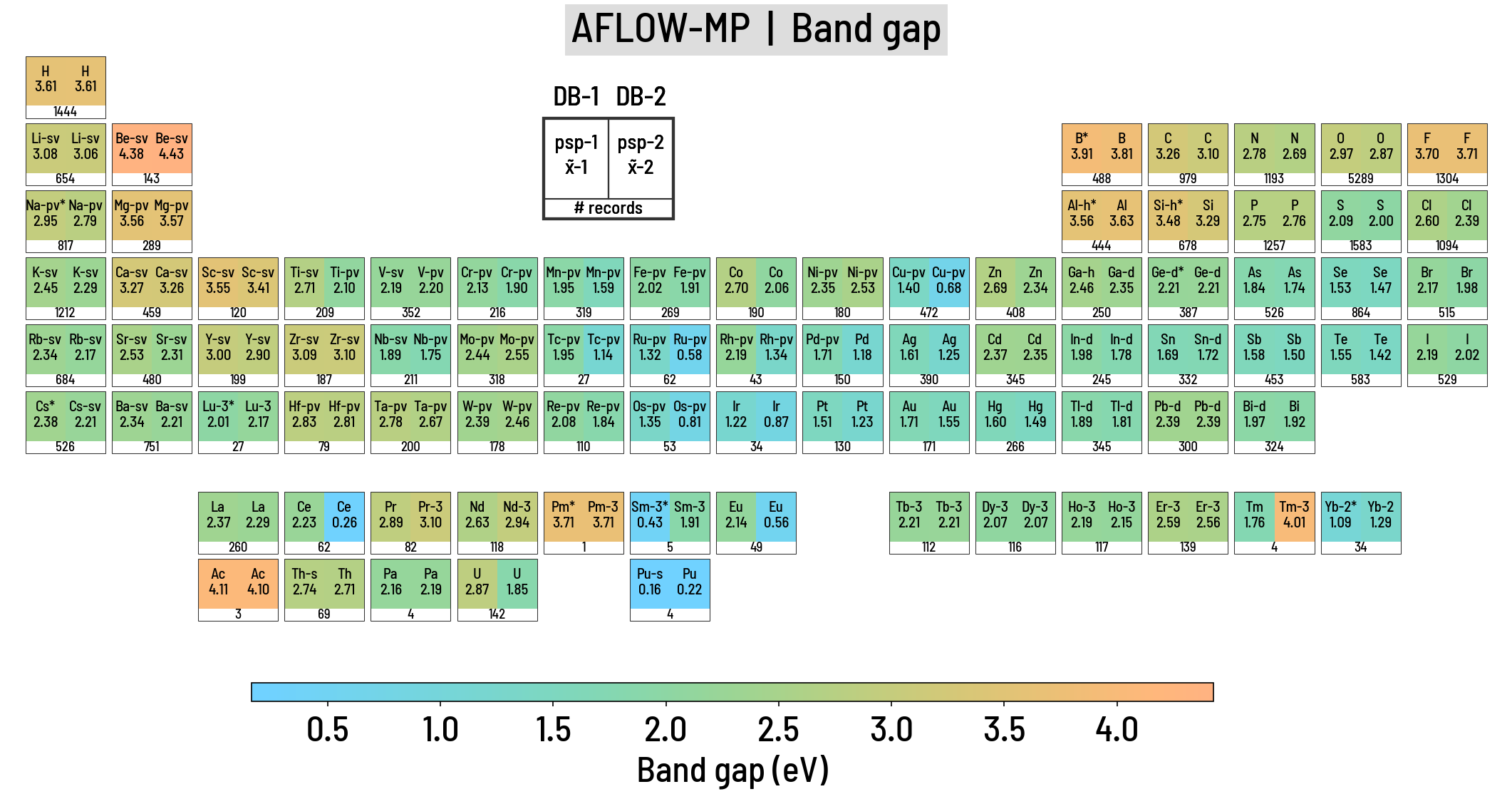}
  \caption{Median values of band gap for compounds containing a certain element
    in the periodic table, for a comparison of AFLOW and MP.
    The VASP PAW potential used for each element and the number of records in
    each comparison are indicated (* indicates more than one pseudopotential
    used in the database overall for that element).}\label{fig:AFLOW-MP-bg}
\end{figure}
\begin{figure}[htb]
  \includegraphics[width=1.00\linewidth]{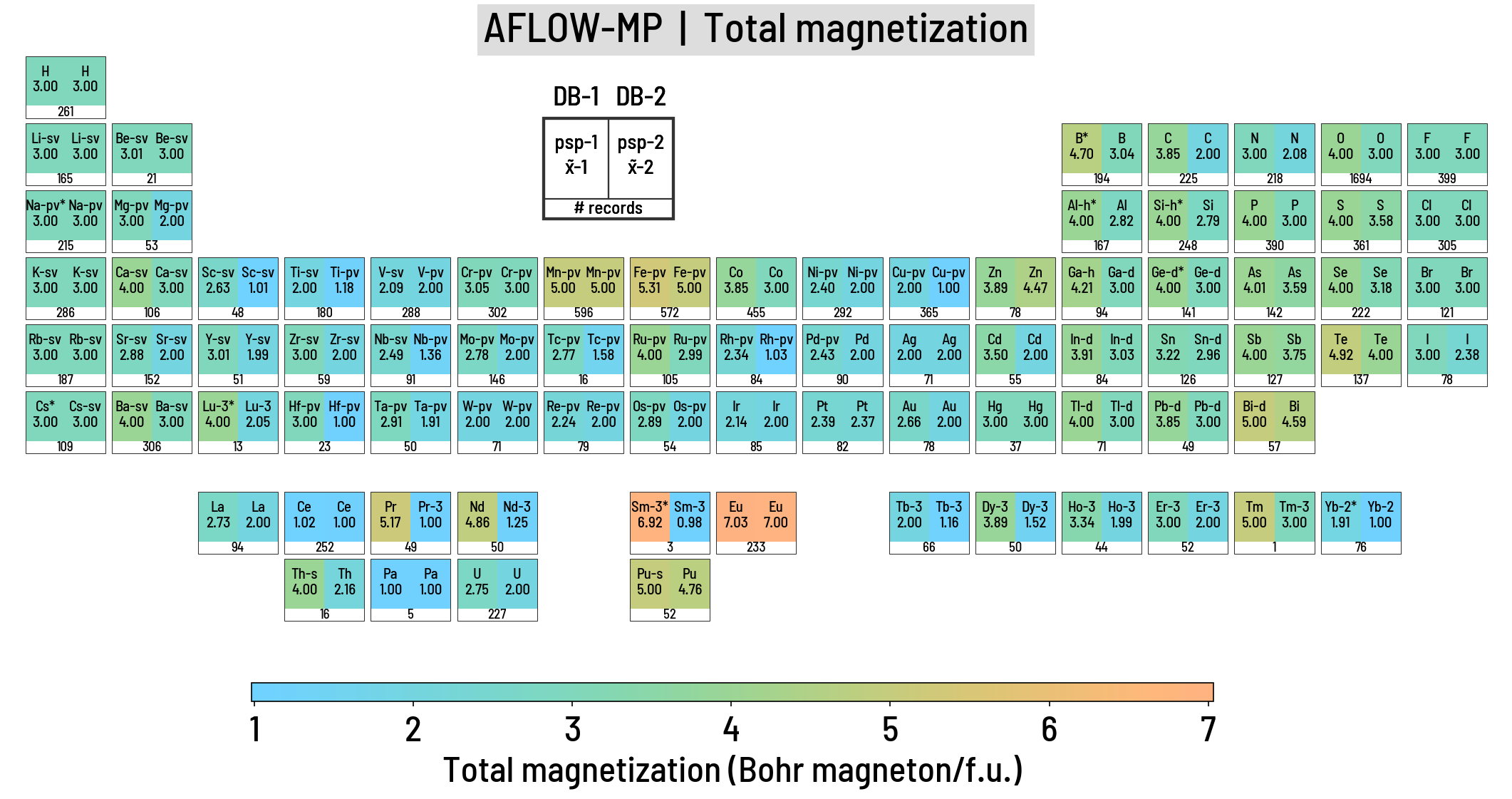}
  \caption{Median values of total magnetization (per formula unit) for
    compounds containing a certain element in the periodic table, for a
    comparison of AFLOW and MP.
    The VASP PAW potential used for each element and the number of records in
    each comparison are indicated (* indicates more than one pseudopotential
    used in the database overall for that element).}\label{fig:AFLOW-MP-mag}
\end{figure}
\begin{figure}[htb]
  \includegraphics[width=1.00\linewidth]{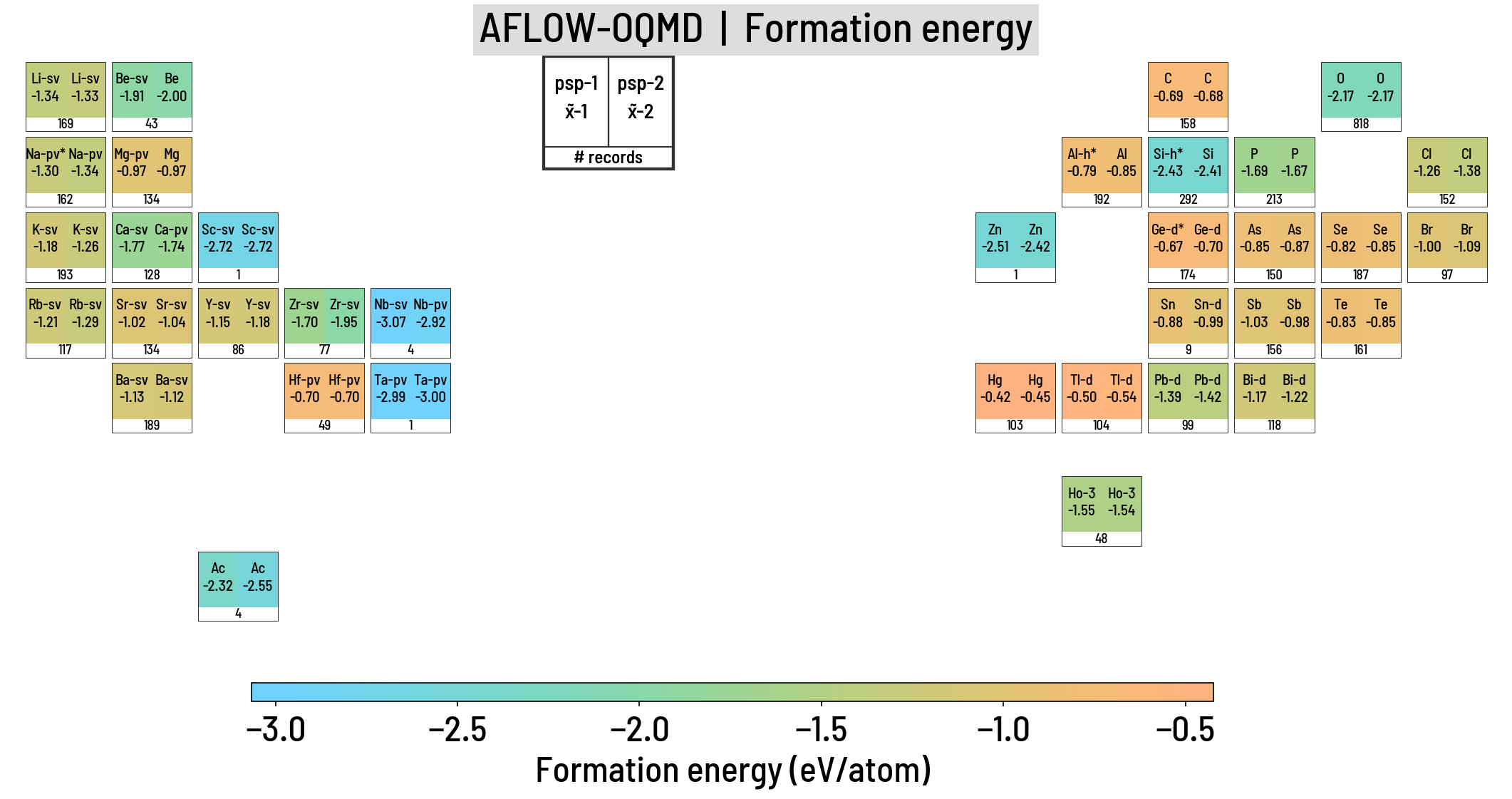}
  \caption{Median values of formation energy for compounds containing a certain
    element in the periodic table, for a comparison of AFLOW and OQMD.
    The VASP PAW potential used for each element and the number of records in
    each comparison are indicated (* indicates more than one pseudopotential
    used in the database overall for that element).}\label{fig:AFLOW-OQMD-dhf}
\end{figure}
\begin{figure}[htb]
  \includegraphics[width=1.00\linewidth]{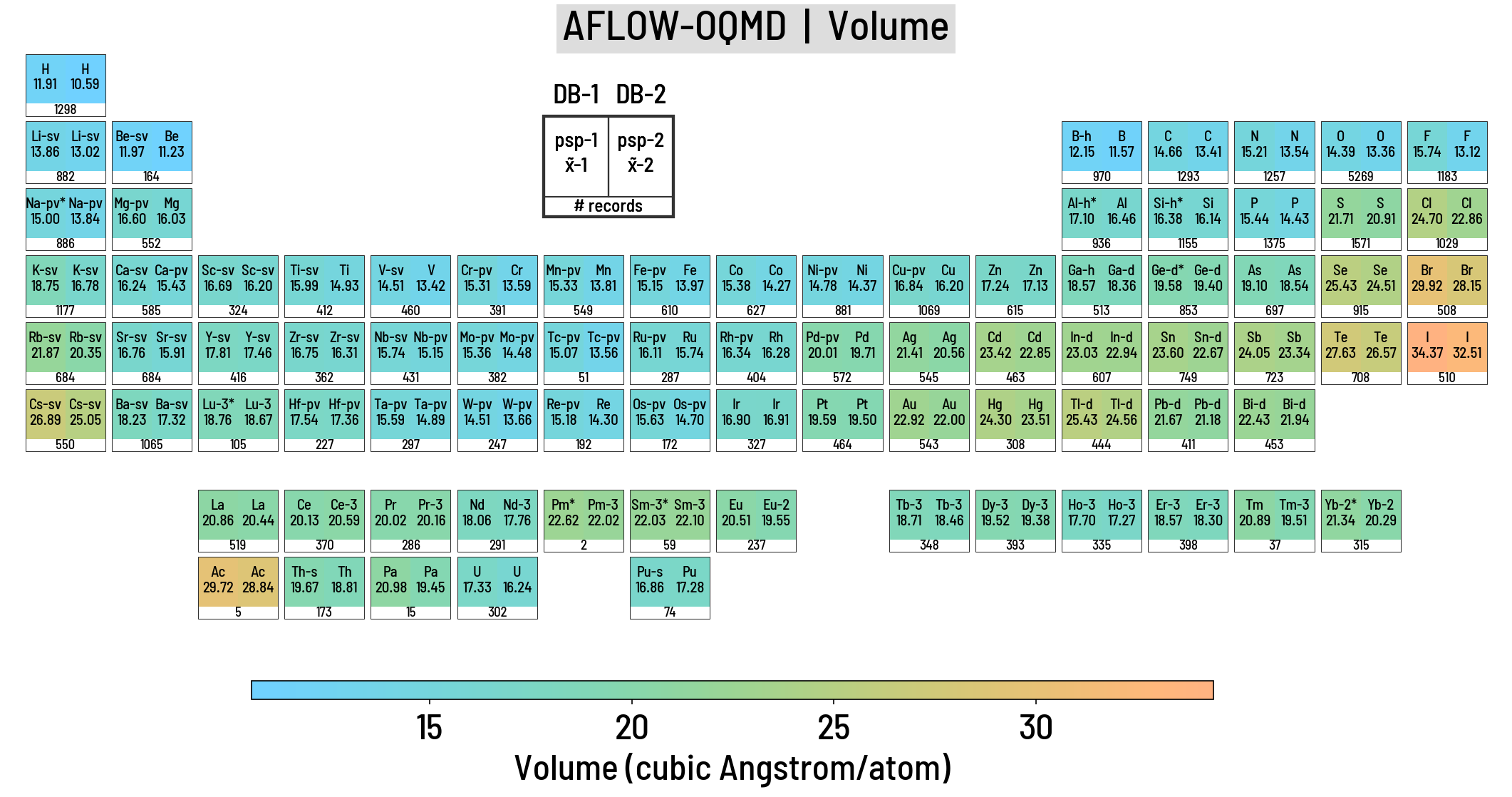}
  \caption{Median values of per-atom volume for compounds containing a certain
    element in the periodic table, for a comparison of AFLOW and OQMD.
    The VASP PAW potential used for each element and the number of records in
    each comparison are indicated (* indicates more than one pseudopotential
    used in the database overall for that element).}\label{fig:AFLOW-OQMD-vol}
\end{figure}
\begin{figure}[htb]
  \includegraphics[width=1.00\linewidth]{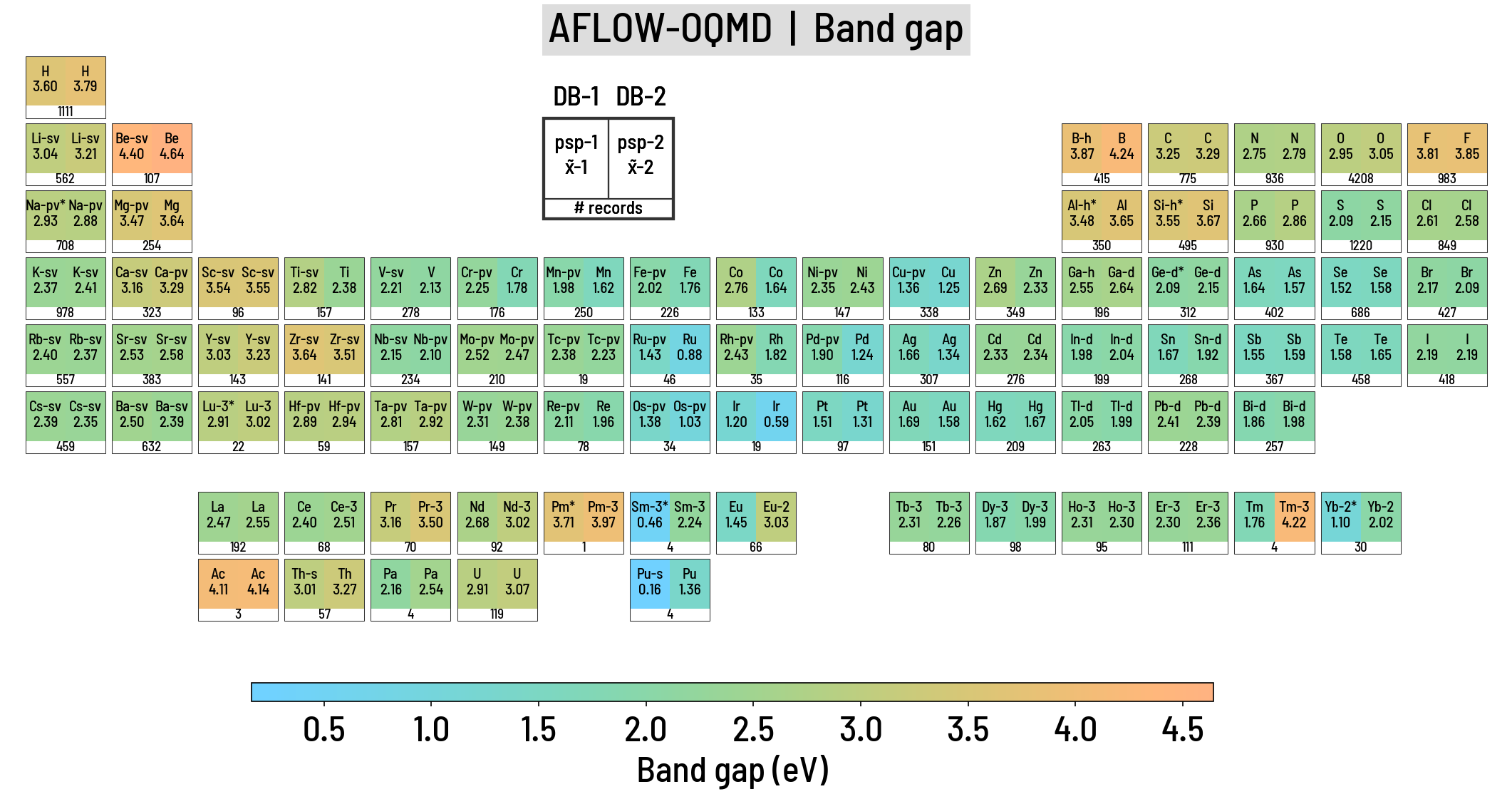}
  \caption{Median values of band gap for compounds containing a certain element
    in the periodic table, for a comparison of AFLOW and OQMD.
    The VASP PAW potential used for each element and the number of records in
    each comparison are indicated (* indicates more than one pseudopotential
    used in the database overall for that element).}\label{fig:AFLOW-OQMD-bg}
\end{figure}
\begin{figure}[htb]
  \includegraphics[width=1.00\linewidth]{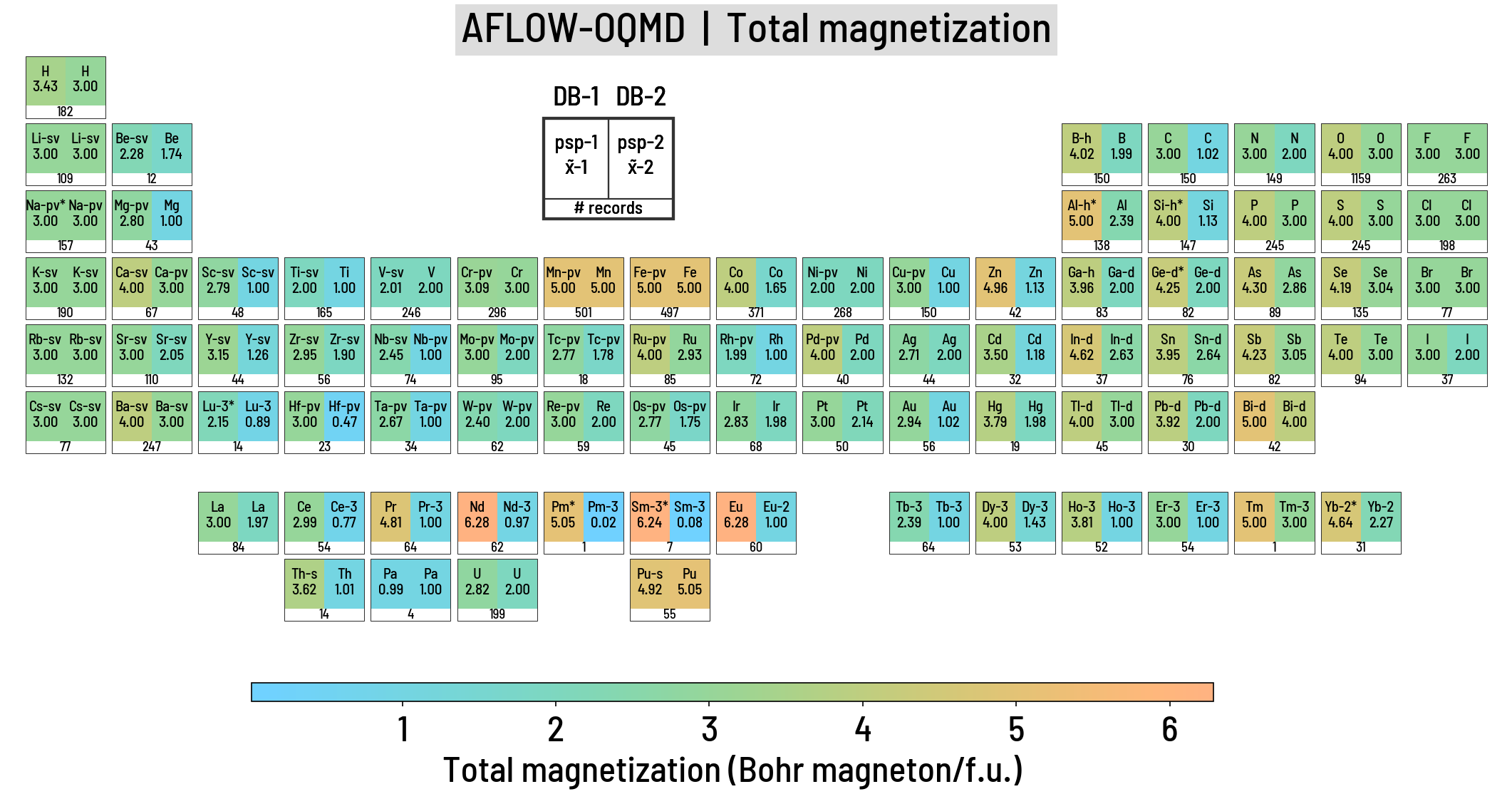}
  \caption{Median values of total magnetization (per formula unit) for
    compounds containing a certain element in the periodic table, for a
    comparison of AFLOW and OQMD.
    The VASP PAW potential used for each element and the number of records in
    each comparison are indicated (* indicates more than one pseudopotential
    used in the database overall for that element).}\label{fig:AFLOW-OQMD-mag}
\end{figure}
\begin{figure}[htb]
  \includegraphics[width=1.00\linewidth]{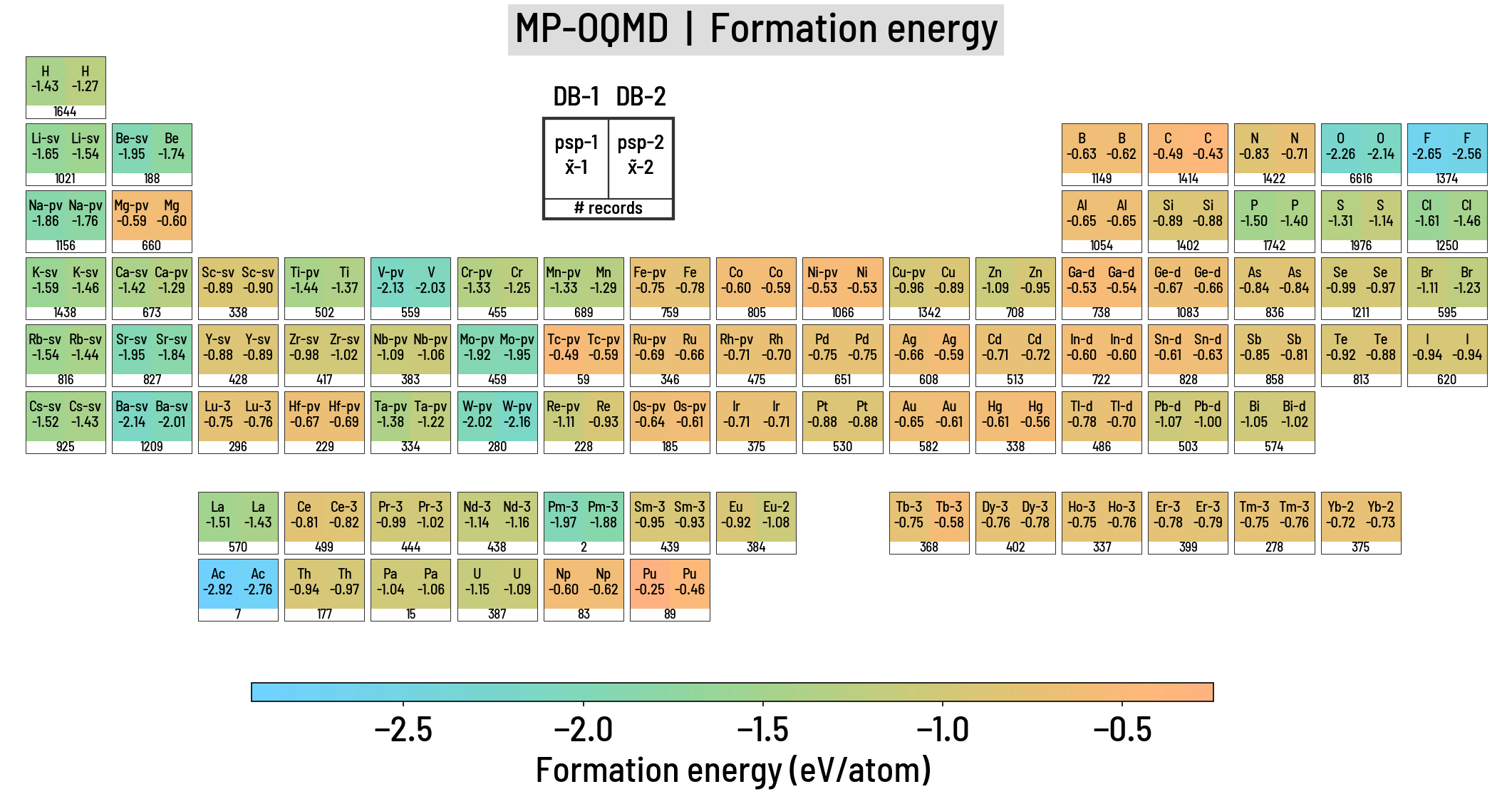}
  \caption{Median values of formation energy for compounds containing a certain
    element in the periodic table, for a comparison of MP and OQMD.
    The VASP PAW potential used for each element and the number of records in
    each comparison are indicated (* indicates more than one pseudopotential
    used in the database overall for that element).}\label{fig:MP-OQMD-dhf}
\end{figure}
\begin{figure}[htb]
  \includegraphics[width=1.00\linewidth]{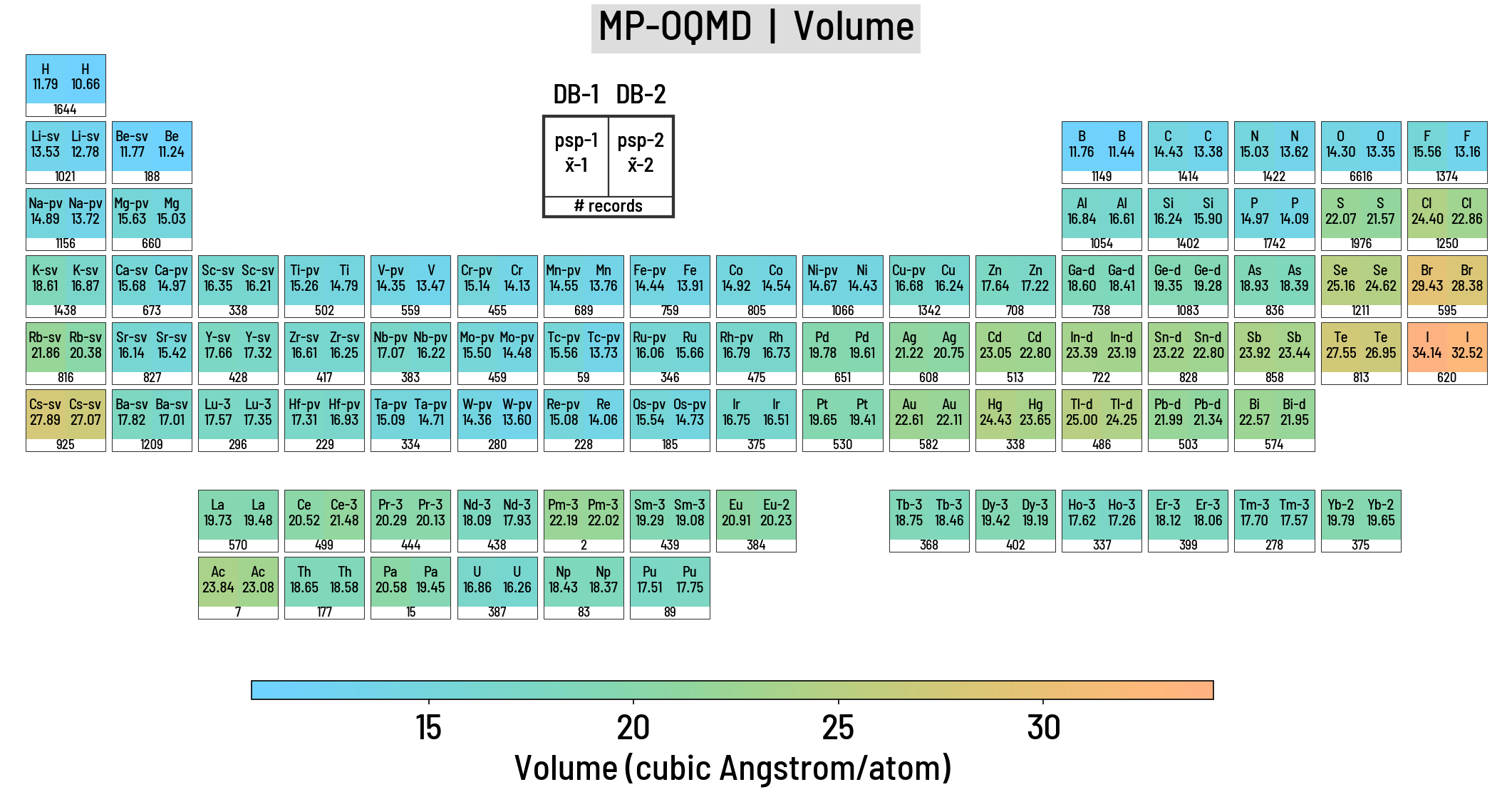}
  \caption{Median values of per-atom volume for compounds containing a certain
    element in the periodic table, for a comparison of MP and OQMD.
    The VASP PAW potential used for each element and the number of records in
    each comparison are indicated (* indicates more than one pseudopotential
    used in the database overall for that element).}\label{fig:MP-OQMD-vol}
\end{figure}
\begin{figure}[htb]
  \includegraphics[width=1.00\linewidth]{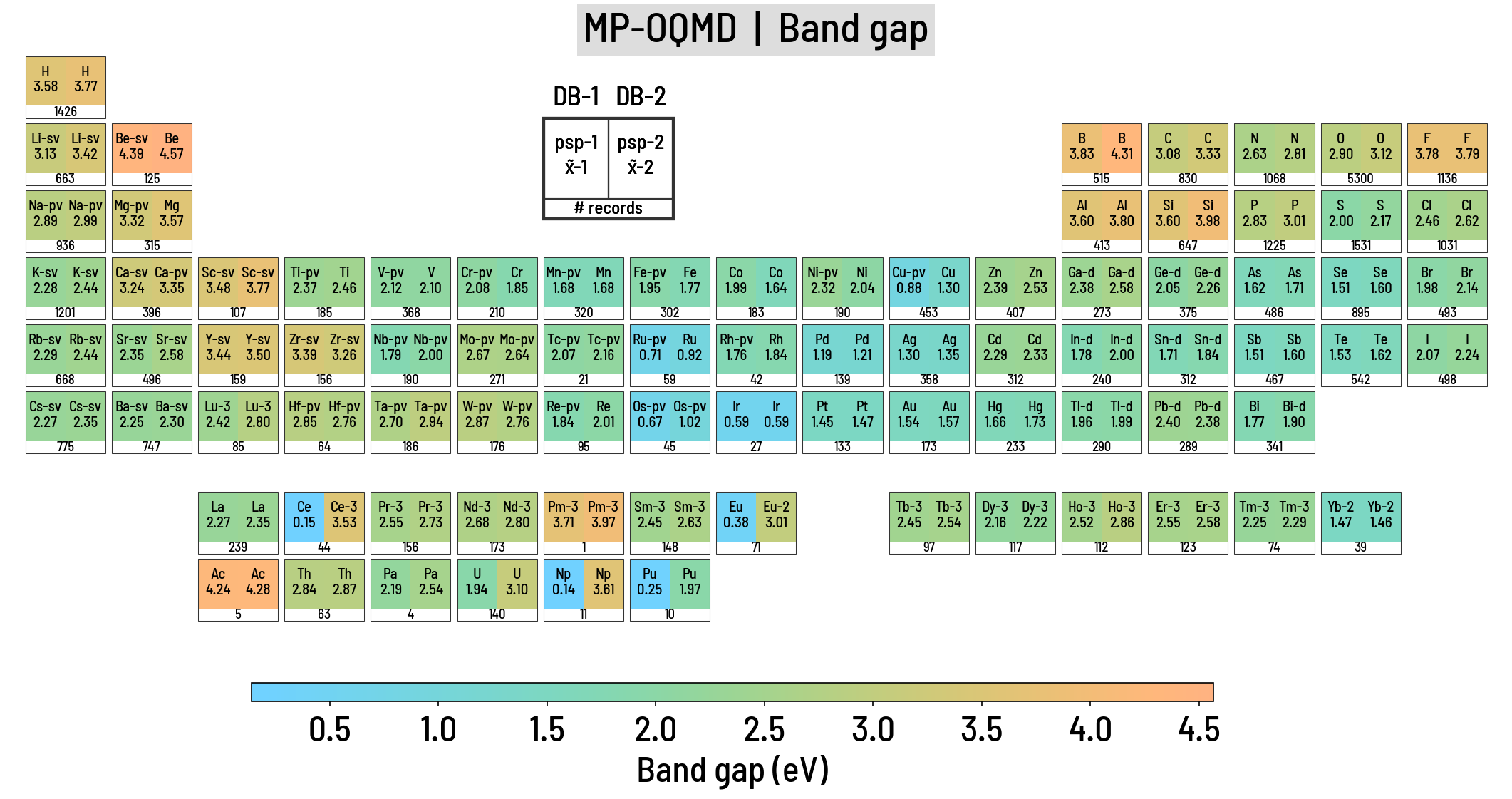}
  \caption{Median values of band gap for compounds containing a certain element
    in the periodic table, for a comparison of MP and OQMD.
    The VASP PAW potential used for each element and the number of records in
    each comparison are indicated (* indicates more than one pseudopotential
    used in the database overall for that element).}\label{fig:MP-OQMD-bg}
\end{figure}
\begin{figure}[htb]
  \includegraphics[width=1.00\linewidth]{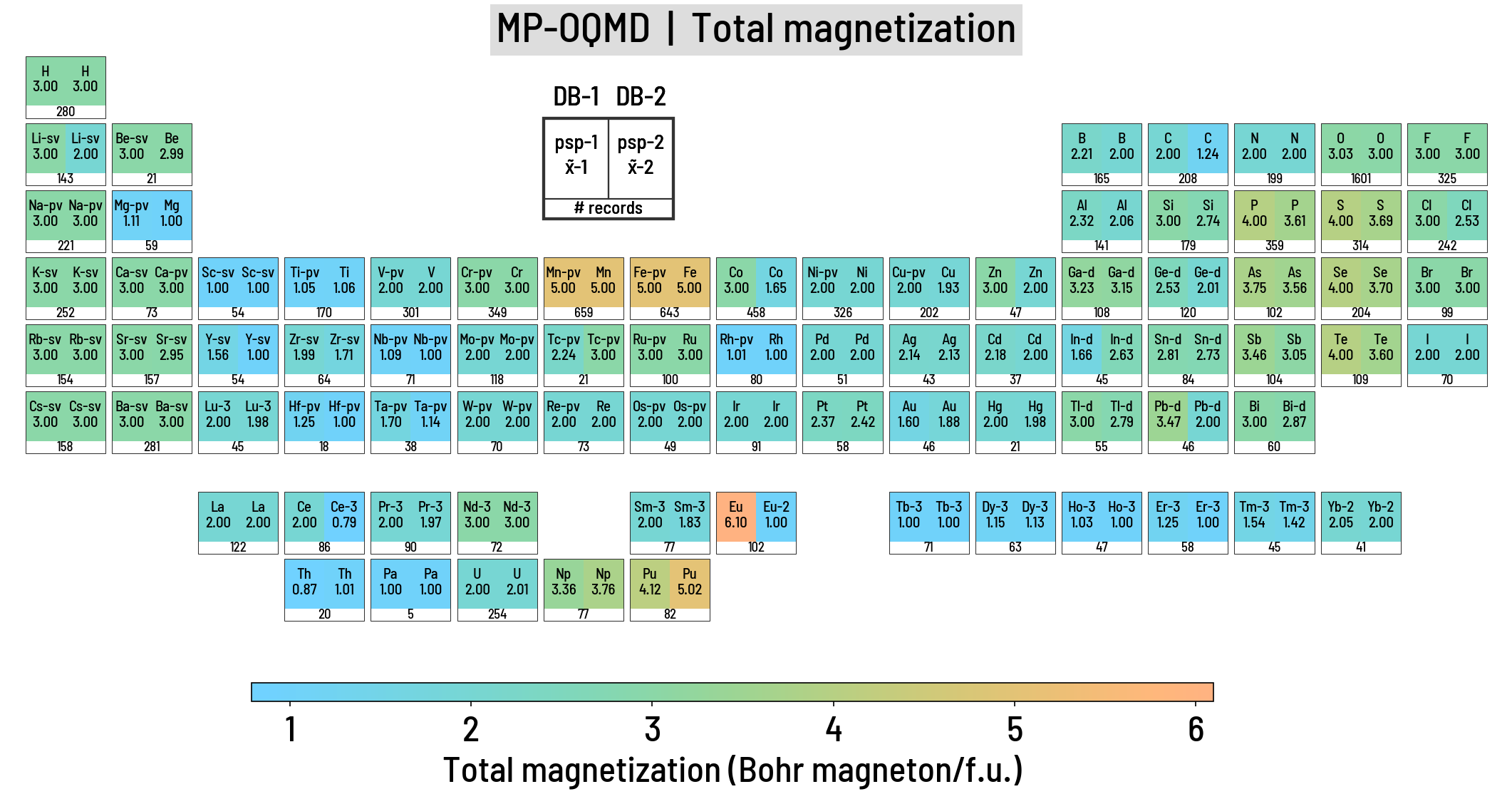}
  \caption{Median values of total magnetization (per formula unit) for
    compounds containing a certain element in the periodic table, for a
    comparison of MP and OQMD.
    The VASP PAW potential used for each element and the number of records in
    each comparison are indicated (* indicates more than one pseudopotential
    used in the database overall for that element).}\label{fig:MP-OQMD-mag}
\end{figure}

To study the source of differences between the various HT-DFT databases, we
collect statistics for the four properties being compared---formation energy,
volume, band gap, total magnetization---averaged over all records containing a
certain element in the periodic table.
For each element, we also present the pseudopotential (psp) used in the two
databases being compared, and the number of comparable records of compounds
containing the element over which statistical quantities are computed.
Note that any missing elemental block in the periodic tables in
Figures~\ref{fig:AFLOW-MP-dhf}--\ref{fig:MP-OQMD-mag} implies that there were
no materials in that comparison set with that element.

%% file: sections/multi_icsd_id_results.tex
\begin{table}[ht]
\caption{The number of records after establishing ICSD ID equivalency for each
property of interest in the AFLOW, Materials Project (MP), and OQMD HT-DFT
databases, as well as for pairwise comparisons of the three
databases.}\label{tbl:prop_table}
\input{multi-icsd-id-tables/records_tally}
\end{table}

\begin{table}[ht]
\caption{Overall statistics (median absolute difference (MAD), interquartile
  range (IQR), Pearson's linear correlation coefficient ($r$), and Spearman's
  rank correlation coefficient ($\rho$)) for the comparison of properties
  across HT-DFT databases.
For each property, records overlapping across a pair of databases are compared
(* for band gap and magnetization, only non-zero values are compared).
Generally, lower MAD, lower IQR, higher $r$, and higher $\rho$ values indicate
better reproducibility of calculated properties.}\label{tbl:summary_table}
\input{multi-icsd-id-tables/corr_coeffs}
\end{table}

\begin{figure}[htp]
\includegraphics[height=0.9\textheight]{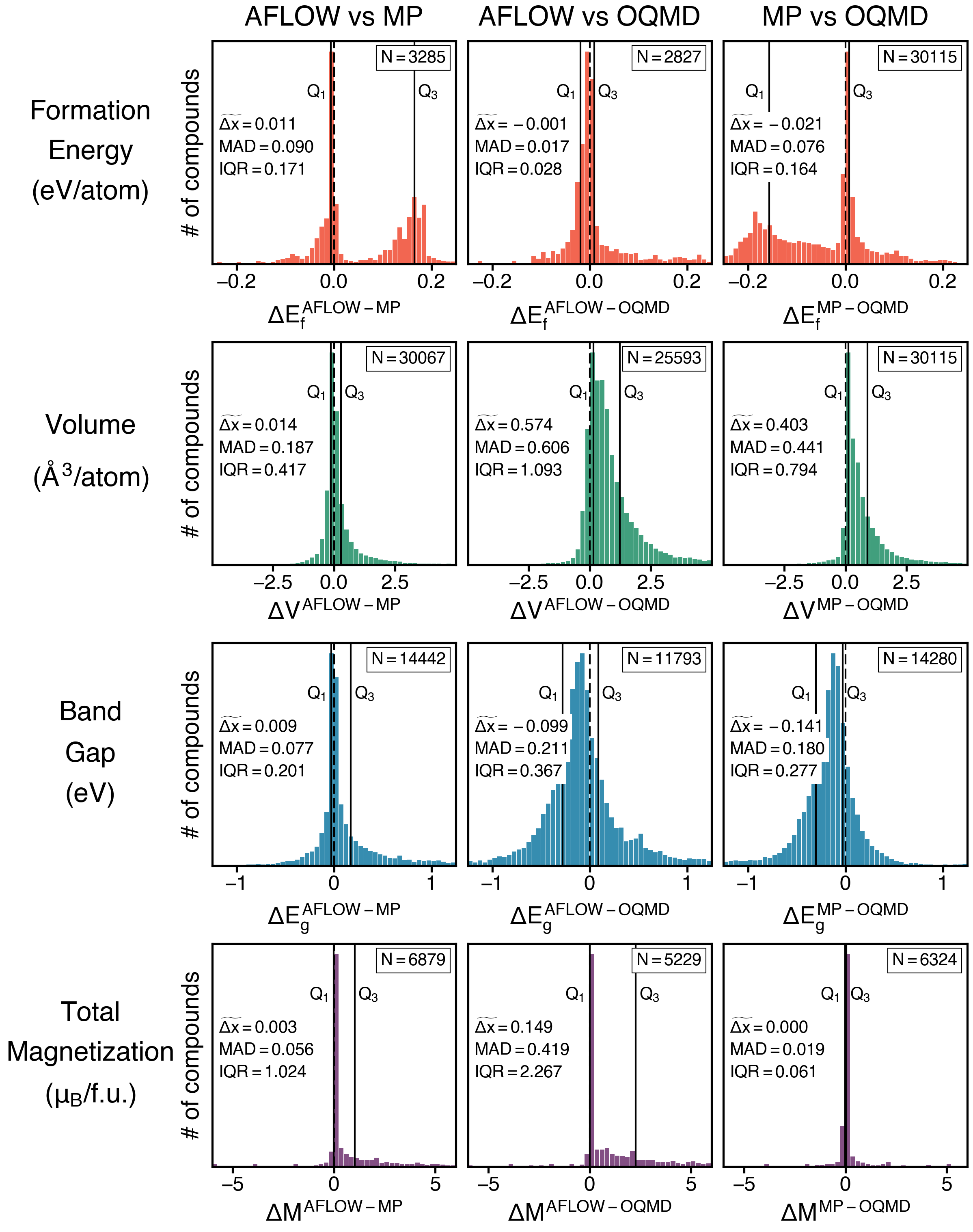}
\caption{Distribution of the differences in calculated properties across HT-DFT
  databases.
Each panel corresponds to a property and pair of databases being compared.
Solid vertical black lines correspond to the first (Q$_1$) and third (Q$_3$)
quartiles of the distribution.
The number of records overlapping across the two databases is shown in the top
right corner of each panel;
the median of distribution ($\widetilde{\Delta x}$), the median absolute
difference (MAD), and the interquartile range (IQR) are noted on the
left.
Note that this figure represents data from the larger comparison datasets
obtained via the structure matching algorithm described in
Section~\ref{sssec:multi_icsd_id}.
}\label{fig:multi_pair_histograms}
\end{figure}

\begin{figure}[htp]
\includegraphics[height=0.85\textheight]{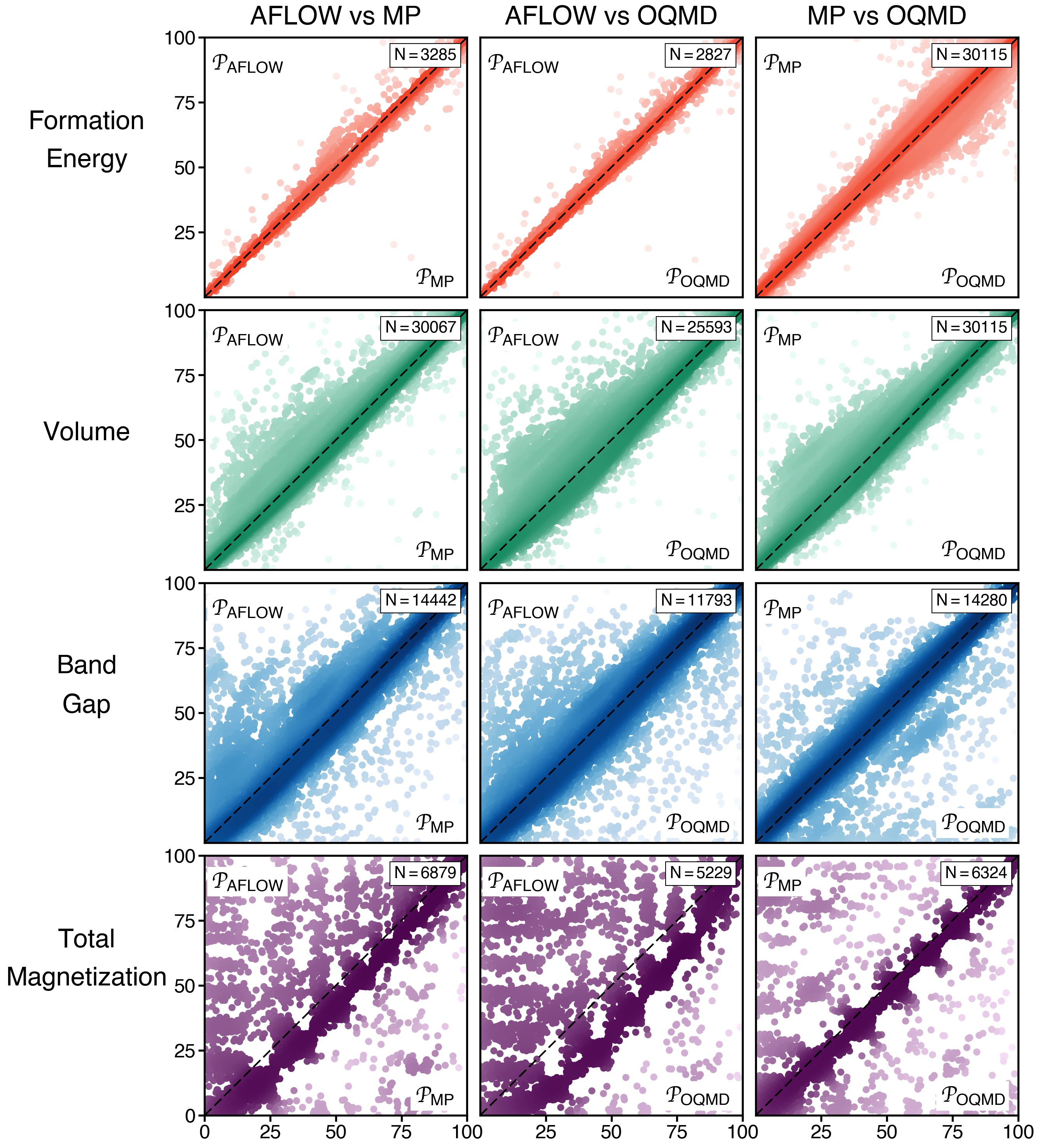}
\caption{Comparison of the calculated properties (formation energy, volume,
  band gap, and total magnetization) over records overlapping across pairwise
  combinations of HT-DFT databases plotted as a percentile rank.
Overall, formation energies and volumes show better reproducibility
than band gaps and magnetizations. The clusters seen in the magnetization
comparisons correspond to nominally integer values of magnetic
moments.
Note that this figure represents data from the larger comparison datasets
obtained via the structure matching algorithm described in
Section~\ref{sssec:multi_icsd_id}.
}\label{fig:multi_pair_percentiles}
\end{figure}

\begin{figure}
\includegraphics[height=0.9\textheight]{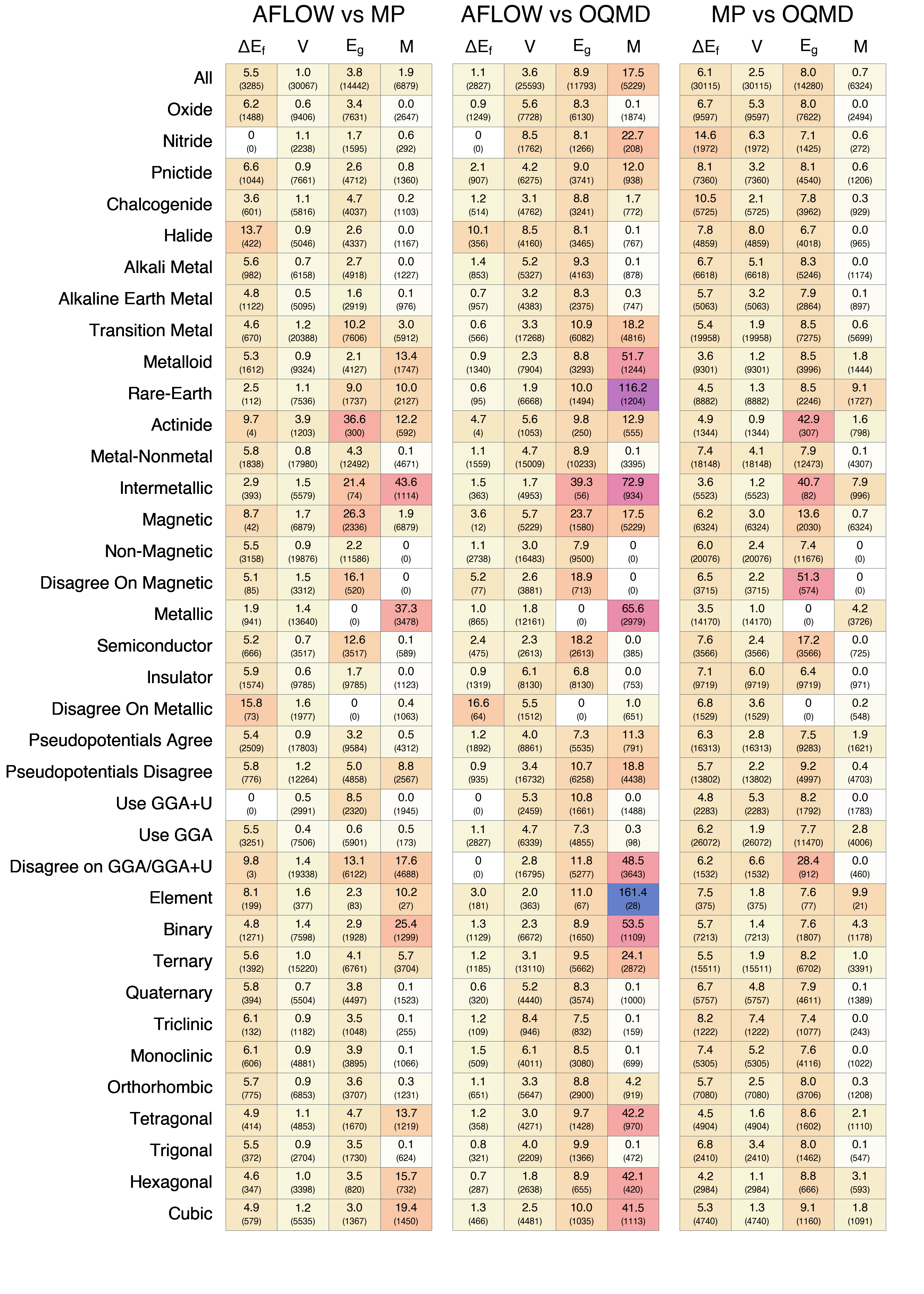}
\caption{Median percent absolute differences between properties (formation
  energy, volume, band gap, total magnetization) calculated in the three
  databases (AFLOW, MP, OQMD), compared two at a time, across various classes
  of materials as defined in Table~III of the main text.
The numbers in parentheses indicate the number of overlapping records belonging
to the respective material class for a given pair of databases.
Trivial comparisons are left blank (e.g., the difference in total magnetization
for non-magnetic compounds).
Note that this figure represents data from the larger comparison datasets
obtained via the structure matching algorithm described in
Section~\ref{sssec:multi_icsd_id}.
}\label{fig:multi_materials_classes}
\end{figure}

\begin{figure*}[ht]
\includegraphics[height=0.9\textheight]{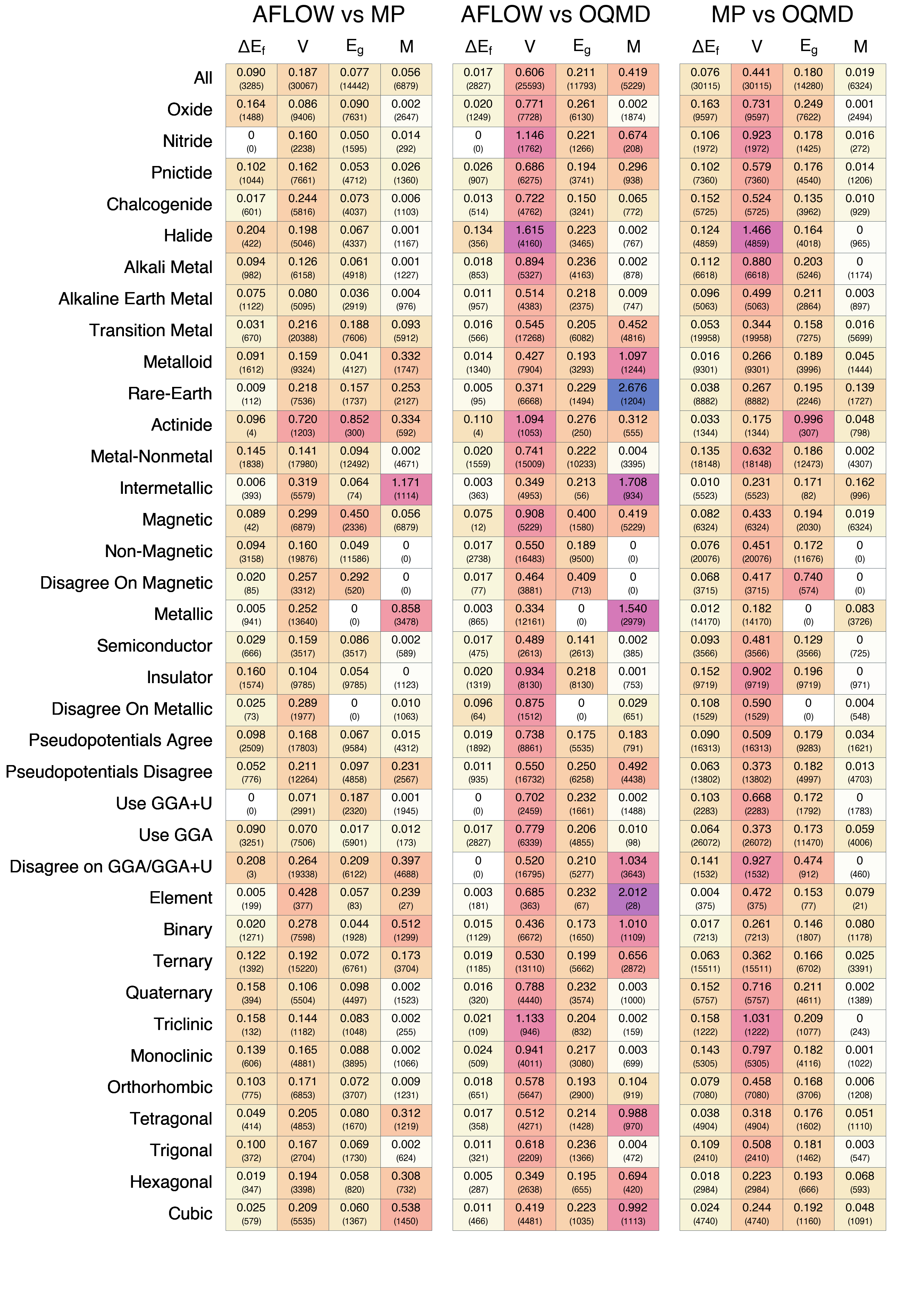}
\caption{Median absolute differences between properties (formation energy,
  volume, band gap, total magnetization are in units of eV/atom,
  {\AA}$^3$/atom, eV, and $\mu_{\rm B}$/formula unit, respectively) calculated
  in the three databases (AFLOW, MP, OQMD), compared pairwise, across various
  classes of materials as defined in Table~III of the main text.
  The numbers in parentheses indicate the number of overlapping records
  belonging to the respective material class for a given pair of databases.
  Trivial comparisons are left blank (e.g., the difference in total
magnetization for non-magnetic compounds).
Note that this figure represents data from the larger comparison datasets
obtained via the structure matching algorithm described in
Section~\ref{sssec:multi_icsd_id}.
}\label{fig:multi_mat-cls-mad}
\end{figure*}

%% file: multi-icsd-id-tables/records_tally.tex
\begin{tabular}{lcccccc}
\toprule
{} & AFLOW & MP & OQMD & AFLOW-MP & AFLOW-OQMD & MP-OQMD \\
\midrule
Formation Energy & 3411 & 48515 & 33281 & 3285 & 2827 & 30115 \\
Volume & 32738 & 48515 & 33281 & 30067 & 25593 & 30115 \\
Band Gap & 32727 & 48515 & 33141 & 30059 & 25466 & 29979 \\
Total Magnetization & 32738 & 48515 & 33281 & 30067 & 25593 & 30115 \\
\bottomrule
\end{tabular}

%% file: multi-icsd-id-tables/corr_coeffs.tex
\begin{tabular}{lcccc@{\hskip 12pt}cccc@{\hskip 12pt}cccc}
\toprule
{} & \multicolumn{4}{c}{AFLOW-MP} & \multicolumn{4}{c}{AFLOW-OQMD} & \multicolumn{4}{c}{MP-OQMD} \\
\midrule
{} & MAD & IQR & $r$ & $\rho$ & MAD & IQR & $r$ & $\rho$ & MAD & IQR & $r$ & $\rho$ \\
\midrule
Formation Energy (eV/atom) &  0.090  &  0.171  &  0.99  &  0.99  &  0.017  &  0.028  &  0.99  &  0.99  &  0.076  &  0.164  &  0.99  &  0.99  \\
Volume (\AA$^3$/atom) &  0.187  &  0.417  &  0.98  &  0.99  &  0.606  &  1.093  &  0.97  &  0.97  &  0.441  &  0.794  &  0.98  &  0.98  \\
Band Gap (eV)* &  0.077  &  0.201  &  0.94  &  0.92  &  0.211  &  0.367  &  0.93  &  0.92  &  0.180  &  0.277  &  0.94  &  0.93  \\
Total Magnetization ($\mu_{\rm B}$/f.u.)* &  0.056  &  1.024  &  0.76  &  0.74  &  0.419  &  2.267  &  0.67  &  0.56  &  0.019  &  0.061  &  0.83  &  0.74  \\
\bottomrule
\end{tabular}

%% file: sections/example_pifs.tex
Example (serialized) PIFs with data queried from each of the three HT-DFT
databases are shown below.

\subsection{Example PIF with queried AFLOW data in the JSON format}
\begin{verbatim}
{
  "category": "system.chemical", 
  "tags": [
    "AFLOW", 
    "ICSD"
  ], 
  "ids": [
    {
      "name": "AUID", 
      "value": "aflow:beda768ce32fec75"
    }, 
    {
      "name": "ICSD", 
      "value": 409808
    }
  ], 
  "references": [
    {
      "url": "http://aflow.org/material.php?id=aflow:beda768ce32fec75"
    }, 
    {
      "url": "http://aflowlib.duke.edu/AFLOWDATA/ICSD_WEB/FCC/Cl6Cs2Mo1_ICSD_409808"
    }, 
    {
      "doi": "http://dx.doi.org/10.1016/j.commatsci.2012.02.002"
    }
  ], 
  "chemicalFormula": "Cl6 Cs2 Mo1", 
  "properties": [
    {
      "scalars": "aflow:beda768ce32fec75", 
      "name": "auid"
    }, 
    {
      "scalars": "aflowlib.duke.edu:AFLOWDATA/ICSD_WEB/FCC/Cl6Cs2Mo1_ICSD_409808", 
      "name": "aurl"
    }, 
    {
      "scalars": "Cl6Cs2Mo1", 
      "name": "compound"
    }, 
    {
      "scalars": "6,2,1", 
      "name": "composition"
    }, 
    {
      "scalars": "9", 
      "name": "natoms"
    }, 
    {
      "scalars": "PAW_PBE", 
      "name": "dft_type"
    }, 
    {
      "scalars": "-34.4871", 
      "units": "eV", 
      "name": "energy_cell"
    }, 
    {
      "scalars": "-3.83189", 
      "units": "eV/atom", 
      "name": "energy_atom"
    }, 
    {
      "scalars": "299.219", 
      "units": "$\\AA^3$", 
      "name": "volume_cell"
    }, 
    {
      "scalars": "33.2465", 
      "units": "$\\AA^3$/atom", 
      "name": "volume_atom"
    }, 
    {
      "scalars": "0", 
      "units": "eV", 
      "name": "Egap"
    }, 
    {
      "scalars": "10,10,10;11,11,11;\\Gamma-X,X-W,W-K,K-\\Gamma,\\Gamma-L,L-U,U-W,W-L,L-K,U-X;20", 
      "name": "kpoints"
    }, 
    {
      "scalars": "1.99872", 
      "units": "$\\mu_B$", 
      "name": "spin_cell"
    }, 
    {
      "scalars": "0.222081", 
      "units": "$\\mu_B$/atom", 
      "name": "spin_atom"
    }, 
    {
      "scalars": "Fm-3m #225,Fm-3m #225,Fm-3m #225", 
      "name": "sg"
    }, 
    {
      "scalars": "Fm-3m #225,Fm-3m #225,Fm-3m #225", 
      "name": "sg2"
    }, 
    {
      "scalars": "225", 
      "name": "spacegroup_orig"
    }, 
    {
      "scalars": "225", 
      "name": "spacegroup_relax"
    }, 
    {
      "scalars": "cubic", 
      "name": "lattice_system_orig"
    }, 
    {
      "scalars": "cubic", 
      "name": "lattice_system_relax"
    }, 
    {
      "scalars": "Cl,Cs,Mo", 
      "name": "species"
    }, 
    {
      "scalars": "Cl,Cs_sv,Mo_pv", 
      "name": "species_pp"
    }, 
    {
      "scalars": "Cl:PAW_PBE:17Jan2003,Cs_sv:PAW_PBE:08Apr2002,Mo_pv:PAW_PBE:08Apr2002", 
      "name": "species_pp_version"
    }, 
    {
      "scalars": true, 
      "name": "is_hubbard"
    }, 
    {
      "scalars": "{\"Cl\": \"0\", \"Cs\": \"0\", \"Mo\": \"2.4\"}", 
      "name": "hubbards"
    }, 
    {
      "scalars": "Cs2MoCl6", 
      "name": "pretty_formula*"
    }, 
    {
      "scalars": [
        "Cl", 
        "Cs_sv", 
        "Mo_pv"
      ], 
      "name": "potentials*", 
      "tags": [
        "PAW", 
        "PBE"
      ]
    }
  ]
} 
\end{verbatim}

\subsection{Example PIF with queried MP data in the JSON format}
\begin{verbatim}
{
  "tags": [
    "Materials Project",
    "ICSD"
  ],
  "references": [
    {
      "url": "https://materialsproject.org/materials/mp-540537"
    },
    {
      "doi": "http://dx.doi.org/10.1063/1.4812323"
    }
  ],
  "ids": [
    {
      "name": "Material",
      "value": "mp-540537"
    },
    {
      "name": "ICSD",
      "value": 1
    }
  ],
  "properties": [
    {
      "name": "last_updated",
      "scalars": "2019-11-14 16:07:39.266000"
    },
    {
      "name": "material_id",
      "scalars": "mp-540537"
    },
    {
      "name": "original_task_id",
      "scalars": "mp-540537"
    },
    {
      "name": "icsd_ids",
      "scalars": [
        1
      ]
    },
    {
      "name": "pretty_formula",
      "scalars": "Cr2Te4O11"
    },
    {
      "name": "final_energy",
      "scalars": -208.54704595,
      "units": "eV"
    },
    {
      "name": "final_energy_per_atom",
      "scalars": -6.133736645588236,
      "units": "eV/atom"
    },
    {
      "name": "volume",
      "scalars": 539.6736877085604,
      "units": "$\\AA^3$"
    },
    {
      "name": "nsites",
      "scalars": 34
    },
    {
      "name": "formation_energy_per_atom",
      "scalars": -1.7626992724264714,
      "units": "eV/atom"
    },
    {
      "name": "e_above_hull",
      "scalars": 0,
      "units": "eV/atom"
    },
    {
      "name": "band_gap",
      "scalars": 0.5996000000000001,
      "units": "eV"
    },
    {
      "name": "is_hubbard",
      "scalars": true
    },
    {
      "name": "hubbards",
      "scalars": "{\"Cr\": 3.7, \"Te\": 0.0, \"O\": 0.0}"
    },
    {
      "name": "volume_per_atom*",
      "scalars": 15.87275552084001,
      "units": "$\\AA^3$/atom"
    },
    {
      "name": "total_magnetization_per_formula_unit",
      "scalars": 5.9998625,
      "units": "$\\mu_B$/formula unit"
    },
    {
      "name": "total_magnetization_per_atom*",
      "scalars": 0.3529330882352941,
      "units": "$\\mu_B$/atom"
    },
    {
      "name": "unit_cell_formula*",
      "scalars": "Cr4 O22 Te8"
    },
    {
      "name": "reduced_cell_formula*",
      "scalars": "Cr2 O11 Te4"
    },
    {
      "tags": [
        "PBE",
        "paw"
      ],
      "name": "potentials*",
      "scalars": [
        "Cr_pv",
        "Te",
        "O"
      ]
    },
    {
      "name": "spacegroup_number",
      "scalars": 14
    },
    {
      "name": "spacegroup_symbol",
      "scalars": "P2_1/c"
    },
    {
      "name": "crystal_system",
      "scalars": "monoclinic"
    },
    {
      "name": "point_group",
      "scalars": "2/m"
    }
  ],
  "category": "system.chemical",
  "chemicalFormula": "Cr2 O11 Te4"
}
\end{verbatim}

\subsection{Example PIF with queried OQMD data in the JSON format}
\begin{verbatim}
{
  "tags": [
    "OQMD v1.2",
    "ICSD"
  ],
  "references": [
    {
      "url": "http://oqmd.org/materials/entry/4189"
    },
    {
      "url": "http://oqmd.org/analysis/calculation/1381320"
    },
    {
      "doi": "10.1007/s11837-013-0755-4"
    },
    {
      "doi": "10.1038/npjcompumats.2015.10"
    }
  ],
  "ids": [
    {
      "name": "Entry",
      "value": 4189
    },
    {
      "name": "ICSD",
      "value": 23036
    },
    {
      "name": "Calculation",
      "value": 1381320
    },
    {
      "name": "FormationEnergy",
      "value": 5636096
    }
  ],
  "properties": [
    {
      "name": "formation__id",
      "scalars": 5636096
    },
    {
      "name": "icsd_id*",
      "scalars": 23036
    },
    {
      "name": "entry__id",
      "scalars": 4189
    },
    {
      "name": "entry__path",
      "scalars": "/home/oqmd/libraries/icsd/23036"
    },
    {
      "name": "entry__duplicate_of__id",
      "scalars": 4189
    },
    {
      "name": "entry__duplicate_of__path",
      "scalars": "/home/oqmd/libraries/icsd/23036"
    },
    {
      "name": "entry__composition__formula",
      "scalars": "Ni5 Si3 Y1"
    },
    {
      "name": "pretty_formula*",
      "scalars": "YSi3Ni5"
    },
    {
      "name": "calculation__id",
      "scalars": 1381320
    },
    {
      "name": "calculation__label",
      "scalars": "static"
    },
    {
      "name": "calculation__configuration",
      "scalars": "static"
    },
    {
      "name": "calculation__band_gap",
      "scalars": 0.0,
      "units": "eV"
    },
    {
      "name": "calculation__energy",
      "scalars": -228.33299659,
      "units": "eV"
    },
    {
      "name": "calculation__energy_pa",
      "scalars": -6.34258323861111,
      "units": "eV/atom"
    },
    {
      "name": "calculation__magmom_pa",
      "scalars": -1.72333333333333e-05,
      "units": "$\\mu_B$/atom"
    },
    {
      "name": "calculation__converged",
      "scalars": true
    },
    {
      "name": "calculation__output__id",
      "scalars": 2069016
    },
    {
      "name": "calculation__output__volume",
      "scalars": 467.937,
      "units": "$\\AA^3$"
    },
    {
      "name": "calculation__output__volume_pa",
      "scalars": 12.9982,
      "units": "$\\AA^3$/atom"
    },
    {
      "name": "calculation__output__natoms",
      "scalars": 36
    },
    {
      "name": "calculation__output__spacegroup__number",
      "scalars": 62
    },
    {
      "name": "formation__delta_e",
      "scalars": -0.7234503675,
      "units": "eV/atom"
    },
    {
      "tags": [
        "v5.2",
        "PAW"
      ],
      "name": "potentials*",
      "scalars": [
        "Ni",
        "Si",
        "Y_sv"
      ]
    },
    {
      "name": "crystal_system*",
      "scalars": "Orthorhombic"
    },
    {
      "name": "is_hubbard",
      "scalars": false
    },
    {
      "name": "hubbards",
      "scalars": "{\"Y\": null, \"Si\": null, \"Ni\": null}"
    }
  ],
  "category": "system.chemical",
  "chemicalFormula": "Ni5 Si3 Y1"
}
\end{verbatim}